\documentclass[10pt,leqno]{amsart}

\usepackage{amssymb,amsthm,amsmath}
\usepackage{graphicx,xcolor,paralist,tikz,fancyhdr,etoolbox,sidecap,subcaption}
\usepackage[hidelinks]{hyperref}
\usepackage[hmargin=2.5cm,vmargin=2.5cm]{geometry}
\usepackage[foot]{amsaddr}

\makeatletter
\newcommand\xleftrightarrow[2][]{%
  \color{blue}\ext@arrow 9999{\longleftrightarrowfill@}{#1}{\hspace{#2}}}
\newcommand\longleftrightarrowfill@{%
  \arrowfill@\leftarrow\relbar\rightarrow}
\makeatother

\begin{document}
\title[System Reliability Engineering in the Age of Industry 4.0: Challenges and Innovations]{System Reliability Engineering in the Age of Industry 4.0: Challenges and Innovations} 
\author[A. Tordeux]{Antoine Tordeux$^*$}
\address{$^*$ \normalfont Corresponding author -- E-mail address: \texttt{tordeux@uni-wuppertal.de}}
\author[T.M. Julitz]{Tim M. Julitz}
\author[I. Müller]{Isabelle Müller}
\author[Z. Zhang]{Zikai Zhang}
\author[J. Pietruschka  et al.]{Jannis Pietruschka}
\author{Nicola Fricke}
\author{Nadine Schlüter}
\author{Stefan Bracke}
\author{Manuel Löwer\smallskip\medskip\\
\textsf{\tiny University of Wuppertal, School of Mechanical Engineering and Safety Engineering\\
Gaußstraße 19, 42119, Wuppertal, Germany}}

\let\thefootnote\relax

\begin{abstract}
In the era of Industry 4.0, system reliability engineering faces both challenges and opportunities. 
On the one hand, the complexity of cyber-physical systems, the integration of novel numerical technologies, and the handling of large amounts of data pose new difficulties for ensuring system reliability. 
On the other hand, innovations such as AI-driven prognostics, digital twins, and IoT-enabled systems enable the implementation of new methodologies that are transforming reliability engineering. 
Condition-based monitoring and predictive maintenance are examples of key advancements, leveraging real-time sensor data collection and AI to predict and prevent equipment failures.
These approaches reduce failures and downtime, lower costs, and extend equipment lifespan and sustainability. 
However, it also brings challenges such as data management, integrating complexity, and the need for fast and accurate models and algorithms.
Overall, the convergence of advanced technologies in Industry 4.0 requires a rethinking of reliability tasks, emphasising adaptability and real-time data processing. 
In this chapter, we propose to review recent innovations in the field, related methods and applications, as well as challenges and barriers that remain to be explored.
In the red lane, we focus on smart manufacturing and automotive engineering applications with sensor-based monitoring and driver assistance systems.
\end{abstract}

\maketitle

\section{Introduction}
\label{sec:1}

The recent advent of computational learning algorithms has revolutionised many disciplines \cite{jordan2015machine}, whether in engineering \cite{jhaveri2022review,vadyala2022review,afshari2023deep}, the natural sciences \cite{carleo2019machine,meuwly2021machine,greener2022guide,sapoval2022current} or the social sciences \cite{vamathevan2019applications,edelmann2020computational,wu2023assessing}. 
This revolution, which began during the 1990s, has also profoundly impacted reliability engineering \cite{duchesne2020recent,xing2020reliability,xu2021machine,afshari2022machine}.
Rather than the emergence of new methods, the rise of Internet-of-Things (IoT), cyber-physical systems and Industry 4.0, along with the growing use of (intelligent) sensors, connected and autonomous systems, data storage, and advanced high-performance computational prognostics, are the key drivers of this change \cite{souza2020survey,ccinar2020machine,jimenez2020towards,ferreira2022remaining}. 
In fact, learning approaches have reached maturity and meet the need to address the reliability of increasingly complex systems. 
\smallskip

Condition monitoring, predictive maintenance, and remote support stand out as remarkable advancements, using real-time data and AI to predict and prevent equipment failures \cite{ran2019survey,schwendemann2021survey,jieyang2023systematic}. 
These approaches reduce downtime, lower costs, and extend equipment lifespan, thereby improving system sustainability. 
Other recent progresses concern system design, e.g.\ fault-tolerant, fail-safe, fail-operational system structures and dynamic reconfiguration, especially for safety-critical systems \cite{oszwald2018dynamic,cerrolaza2020multi}. 
In addition, computer-aided reliability analysis based on simulation, digital twins and real-time data assimilation allows, for example, the effects of degradation and other dynamic factors in mechanical and cyber-physical systems to be taken into account in real time   \cite{liu2021review,thelen2022comprehensive,thelen2023comprehensive}.
Applications cover various fields, including (smart) manufacturing and industrial production \cite{ccinar2020machine,aldrini2024fault,friederich2024reliability}, energy \cite{kumar2020reliability,duchesne2020recent}, robotics, transport and logistics \cite{martyushev2023review}, (smart) cities and healthcare \cite{liu2020failure}.
The increasing complexity of systems and algorithms, the integration of new digital technologies and the management of large amounts of data generated in real time remains major challenges. 
Overall, the convergence of advanced technologies in Industry 4.0 requires rethinking reliability tasks with a focus on adaptability and real-time data processing. 
\smallskip

The aim of this chapter is to review the latest innovations, applications and relevant methodologies in the field, as well as the challenges and barriers that remain to be addressed. 
A red lane in the chapter is dedicated to smart manufacturing, automotive and traffic engineering applications, in particular sensor-based monitoring and driver assistance systems.
The chapter is structured as follows.
We identify and review in section~\ref{sec:Innov} three recent innovation domains in advanced system reliability engineering, namely real-time IoT-enabled sensor-based reliability analysis 
in section~\ref{subsec:Monitoring}, improved reliability through system design 
in section~\ref{subsec:Architectur}, and simulation-aided reliability analysis 
in section~\ref{subsec:Simulation}.
Then we present some challenges and discuss barriers that remain to be explored in section~\ref{sec:Challenges}, covering difficulties related to safeguarding AI and real-time data processing in section~\ref{subsec:SafeAI_BigData};  system complexity and component interdependence in section~\ref{subsec:SystComplex}; and human-machine collaboration and sociotechnical systems in section~\ref{subsec:HuamFactor}.
Finally, in section~\ref{sec:Stat}, we review related statistical methods and machine learning techniques for multivariate datasets. 
The methods include unsupervised learning approaches for data visualisation, dimension reduction and clustering in section~\ref{subsec:Unsupervised}, and supervised learning methods for advanced predictive analytics, e.g., multivariate regression, classification, and machine learning techniques in section~\ref{subsec:Supervised}.

\section{Innovations}
\label{sec:Innov}

\begin{figure}[b]
    \centering\vspace{-5mm}
    \footnotesize\input{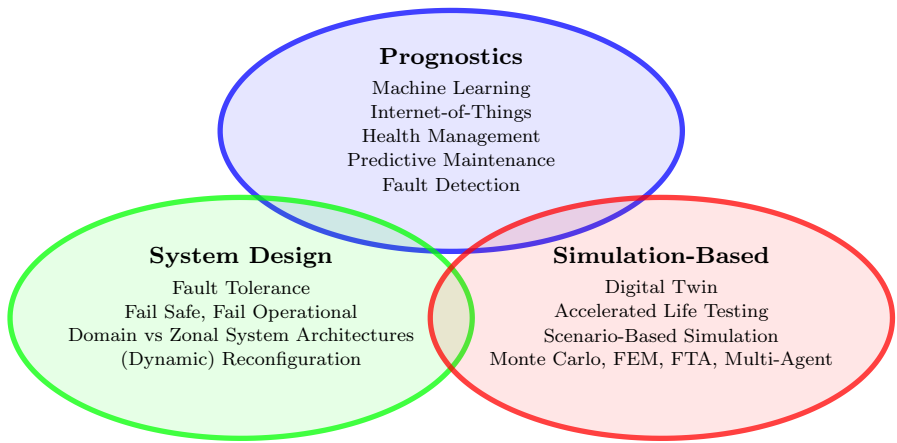}\vspace{-5mm}
    \caption{Main sectors that contribute to advanced reliability engineering techniques during the last decades.}
    \label{fig:MainSector}
\end{figure}
 
\begin{figure}[b]
    \centering
    \input{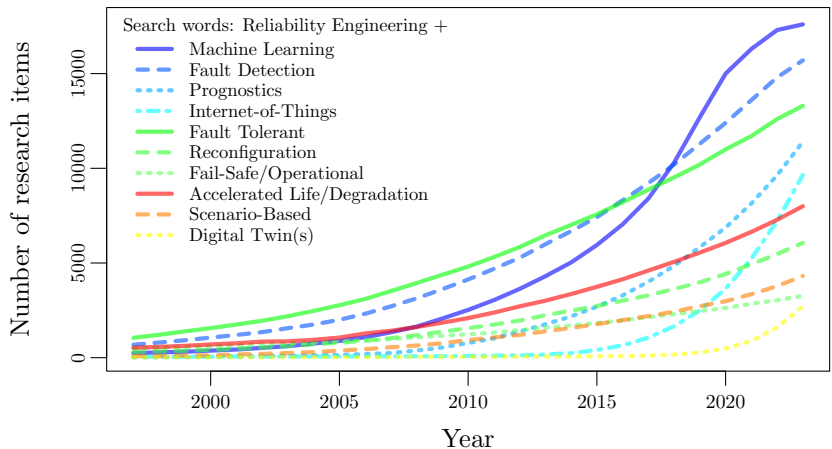}
    \caption{Cumulative number of research articles from 1997 to 2023 with the tags 'Reliability Engineering' and related keywords in Google Scholar (query made on 01.10.2024).}
    \label{fig:NbPublis}
\end{figure}

In this section, we review the most important recent advances in the three innovation areas of prognostics, system design and simulation-based reliability analysis (see Figs.~\ref{fig:MainSector} and \ref{fig:NbPublis}).
These research areas of reliability engineering have benefited from important innovations in recent decades. 
First, sensor-based monitoring based on machine learning techniques, supported by the Internet of Things and Industry 4.0, has shown significant progress in prognostics and health management, real-time condition monitoring, fault detection, self diagnosis and predictive maintenance. 
Then system design, which showed how advanced architectures can improve the system reliability through redundant configurations, fault-tolerance, fail-safe, fail-operational architectures, dynamic reconfiguration and automation. 
Finally important innovations have also been provided by simulation-based reliability analysis, e.g. using digital twins, accelerated life testing, or scenario-based simulation. 
Figure~\ref{fig:NbPublis} shows the number of research articles over the last twenty years with related keywords in the Google Scholar engine. 
The trends are clearly more than linear for all the selected keywords. 
We can observe that the keywords \emph{Digital Twin(s)}, \emph{Internet-of-Things} and, to a lesser extent, \emph{Prognostics} and \emph{Fault Detection} are very topical. 
The keyword \emph{Machine Learning} peaked about five years ago, although it is still the most cited keyword. 
\emph{Fault Tolerant}, \emph{Accelerated Life/Degradation}, \emph{Reconfiguration}, \emph{Fail-Safe/Operational}, and \emph{Scenario-Based} are keywords that were established earlier in the 2000s.

\subsection{Prognostics}
\label{subsec:Monitoring}
Sensor-based monitoring has emerged as an innovation in reliability engineering, driven by advances in data acquisition, real-time analysis, and integration with machine learning (ML) and AI technologies. 
IoT integration in the context of Industry 4.0 enables continuous or periodic monitoring of operating conditions in systems by using for example {\bf Wireless Sensor Networks} (WSN). 
WSN collect vast amounts of data (i.e. vibration, temperature, pressure, etc.) remotely and decentralised. 
For (real-time) analysis purposes, IoT platforms provide access to this transmitted data, where it is processed to assess system health.
The data acquisition is a crucial step in the process. 
Indeed, the labelled data enables machine learning algorithms to be trained and validated. 
Numerous open-access online data repositories exist for this purpose, such as Kaggle with specif keywords\footnote{\href{https://www.kaggle.com/datasets?search=predictive+maintenance}{\texttt{https://www.kaggle.com/datasets?search=predictive+maintenance}}}, NASA Prognostics Data Repository\footnote{\href{https://www.nasa.gov/intelligent-systems-division/discovery-and-systems-health/pcoe/pcoe-data-set-repository/}{\texttt{https://www.nasa.gov/intelligent-systems-division/discovery-and- systems-health/pcoe/pcoe-data-set-repository/}}}, UC Irvine Machine Learning Repository \footnote{\href{https://archive.ics.uci.edu/}{\texttt{https://archive.ics.uci.edu/}}}, etc.

A comprehensive framework that incorporates these innovations is {\bf Prognostics and Health Management} (PHM). 
PHM consists of four main components: 
\begin{enumerate}
    \item Data Acquisition and Condition Monitoring,
    \item Diagnostics,
    \item Prognostics,
    \item Health Management.
\end{enumerate}  
Within the PHM framework, data acquisition and condition monitoring integrates sensor-based monitoring with data analytics and machine learning techniques to monitor system health. 
Such a area relies on advanced sensor technology (e.g., WSN).
Diagnostics aims to detect abnormal system behaviour, identify and isolate faults, and analyse the root causes of system failures, while Prognostics goes a step further by estimate the remaining useful life (RUL) and failure prediction based on current operational conditions. 
System maintenance strategy (i.e. predictive maintenance) and performance optimisation are enabled through the use of diagnostics and prognostics. 
Optimising maintenance schedules by using diagnostics and prognostics minimises unplanned downtime and maximises the operational life of (industrial) systems and performance optimisation, which is referred to Health Management.
Many reviews exist on the topics, see, e.g., \cite{souza2020survey,ccinar2020machine,jimenez2020towards,ferreira2022remaining,ran2019survey,schwendemann2021survey,jieyang2023systematic} and references therein.


\subsubsection{Fault detection and diagnosis}
\label{subsubsec:FaultDetect_Diagnosis}
	
\begin{figure}
    \centering
    \qquad\includegraphics[width=.9\textwidth]{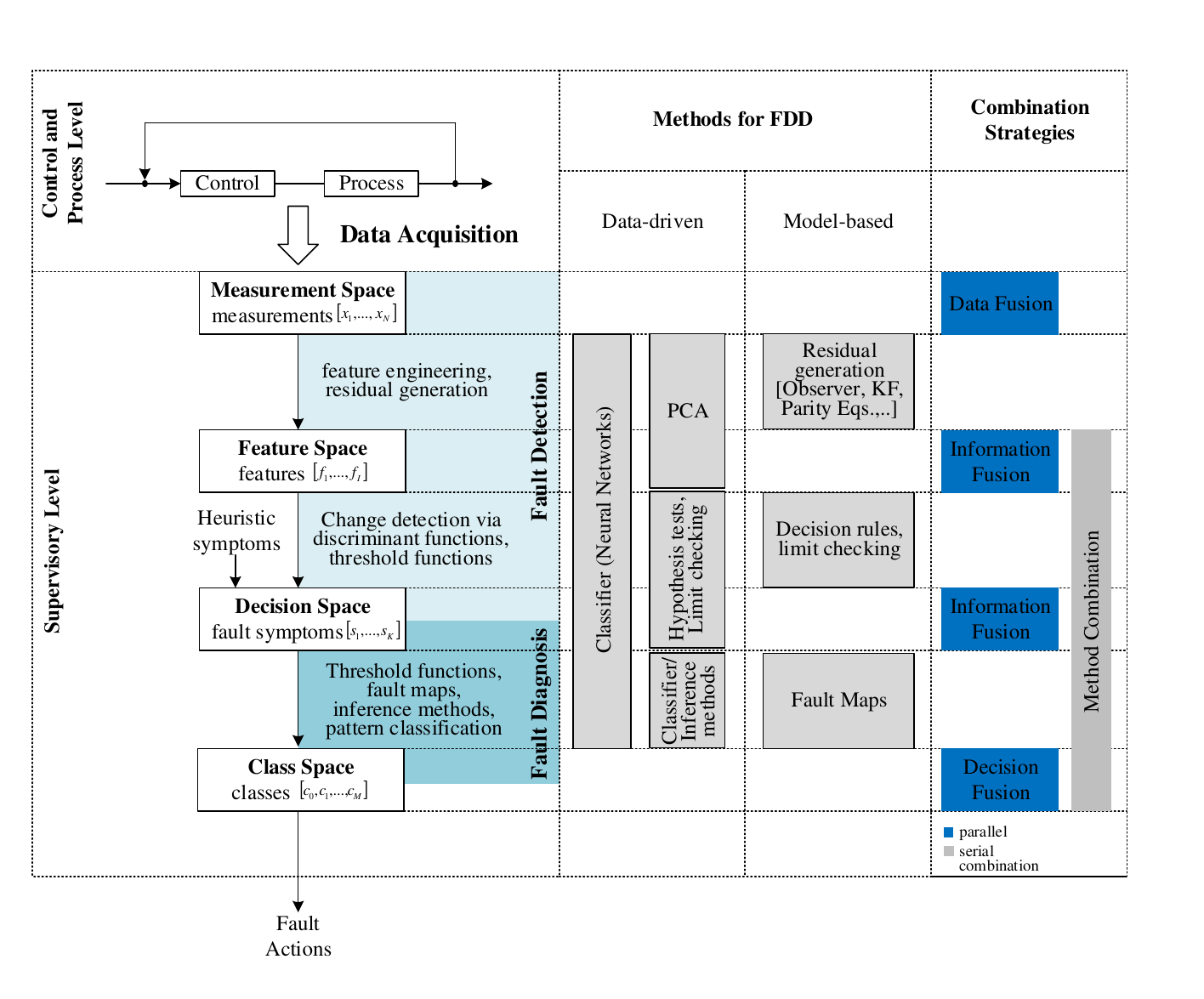}
    \caption{Transformations within the FDD system, used data-driven and model-based methods and combination strategies for hybrid approaches \cite{wilhelm2021overview}}
    \label{fig:workflow_FDD}
\end{figure}

In modern industrial applications where sensor networks are implemented, a large amount of process data is available. They contain important information about process dynamics and system health. In general, the diagnostics process integral to the {\bf Fault Detection and Diagnosis} (FDD) framework is a series of transformations on system measurements \cite{venkatasubramanian2003review}, shown in figure \ref{fig:workflow_FDD}. The FDD system uses these measurements from the process as an input. Then, in the next step features are derived from measurements through feature engineering or residual generation. 
Features act as health indicators of faulty (abnormal) process behaviour and provide a basis for detecting changes, which refer to fault symptoms. For example \cite{zhou2022construction} provides an overview of the health indicator (HI) construction for rotating machinery. In the decision space, changes in the operating conditions are detected with threshold or discriminant functions.  
Subsequently, the fault diagnosis constitutes a mapping between the set of fault symptoms and the set of fault classes. 
In order to ascertain the fault classes, including the normal operating condition, a variety of methods may be employed, such as pattern recognition, classifiers, inference methods or threshold functions.
Afterwards the output of the diagnostics process, i.e. features and fault classes, may be used as an input for prognostics and health management \cite{zio2022prognostics, tchakoua2013review}. 

The methods employed in diagnostics can be broadly classified into two categories: data-driven and model-based, shown in figure \ref{fig:workflow_FDD}. In addition, the work of \cite{wilhelm2021overview} considered knowledge-based approaches. 
Data-driven methods extract information from historical process data and are generally divided into supervised (i.e., classification and regression techniques) and unsupervised methods (i.e., statistical models and clustering techniques).
Supervised methods train algorithms to historical data with the objective of learning and distinguishing patterns within the high-dimensional space of process measurements or features. The patterns may then be used to perform fault detection and fault diagnosis, specifically to predict the fault symptoms or classes for new data. However, such methods require labelled training data, whereby for each input feature vector, the corresponding class label must be known \cite{wilhelm2021overview}.

Unsupervised methods do not require class labels and are therefore suitable for a wider range of applications where these are not available. Most unsupervised methods focus only on the first or second transformation step to extract features or fault symptoms \cite{wilhelm2021overview}. Principle component analysis (PCA) is the most common approach to dimensionality reduction and feature extraction. Typical data-driven methods especially used for classification purposes are artificial neural networks (ANN), support vector machines (SVM), k-nearest-neighbour (kNN), decision trees, and Random Forest, see e.g. \cite{ribeiro2022fault, islam2019reliable, aly2005survey}.

Model-based approaches rely on numerical reproduction in the form of mathematical equations of the real physical process. The mathematical equations are based on the knowledge of the input-output relationships of the process \cite{isermann2006fault}. Due to these known input-output relationships, the model requires less process data compared to data-driven approaches.
The operating state of the process is determined by comparing measurements of the real process state with the state predicted by the physical model. An inconsistent comparison indicates a faulty condition. Specific inconsistencies are referred to as residuals. The residuals correspond to the characteristics shown in the workflow fig.\ref{fig:workflow_FDD}. The subsequent fault diagnosis can be carried out using threshold logic functions, for example. Typical methods for FDD based on physical models are bond graphs, parity equations, state estimation (e.g., using Kalman filter), state observers or simple limit and trend checking, see e.g. \cite{samantaray2006diagnostic,isermann1993fault, dardanelli2010model, frank1997survey, isermann2006fault}.

Hybrid FDD approaches combine two or more established methodologies. Additionally, some approaches consider direct integration of a priori knowledge. Hybrid FDD methods may be combined parallel or serial through a variety of algorithms, see \cite{wilhelm2021overview}. 
In the case of a data set comprising a significant amount of unlabelled data, the following hybrid methods may be employed to automatically expand the data set, for instance:
\begin{itemize}
    \item {\bf Semi-supervised}: A model is trained using a small, manually labelled data set, which is then applied to unseen, unlabelled data. Subsequently, the previously unlabelled data is incorporated into the training data set, with a label selected by the model itself. It is crucial that the base data set with the labelled data is selected to be representative of the use case, in order to facilitate the optimal training of the model. In an iterative process, the data labelled by the model is then added to the training data set, with the objective of optimising the model \cite{Bruggemann.2022, Leva.2022}.
    
    \item {\bf Activ Learning}: The model is trained in an iterative manner, whereby the most "uncertain" or informative data points are selected for human labelling. This approach entails that only the most necessary data is labelled manually, while the model assumes responsibility for the remaining labels automatically. This reduces the labelling effort, but with the potential for incorrect labelling of data points in which the model is certain \cite{Chen.2012, ChristophRosebrock.2021}.
    
\end{itemize}

\subsubsection{Remaining Useful Life Prediction and Sampling Methods}
\label{subsubsec:RULPrediction_Sampling}

Predictive maintenance strategies are generally based on real-time prediction of the {\bf Remaining Useful Life} (RUL) of a system. These strategies allow for proactive maintenance planning, which helps avoid unexpected downtime and optimizes operational efficiency. As RUL predictions reach critical thresholds, maintenance interventions are initiated, enabling actions to be taken before failures occur. RUL prediction methods are widely applicable, particularly in industries that rely on battery and bearing maintenance, as reviewed in \cite{hu2020battery, ge2021review, hasib2021comprehensive, jieyang2023systematic}.

Common predictive maintenance algorithms can be categorised into \emph{shallow} and \emph{deep learning} methods, each providing unique advantages. In shallow learning, methods like \emph{Logistic Regression} are computationally efficient and simple to implement, though they assume linear relationships, which limits their suitability in complex environments \cite{hu2020battery, ge2021review}. \emph{Artificial Neural Networks} (ANNs) are also common in shallow learning; they store complex patterns and are robust to noise but may yield unpredictable results under certain conditions \cite{lei2018machinery}.

Other popular shallow learning techniques include \emph{Support Vector Machines} (SVM), which perform well in high-dimensional data spaces, though they are less effective with large datasets. \emph{Deep Belief Networks} (DBNs) require minimal labeled data and have quick training times but pose challenges in determining optimal network structures \cite{hasib2021comprehensive}. Deep learning approaches, such as \emph{Convolutional Neural Networks} (CNNs) and \emph{Recurrent Neural Networks} (RNNs), are highly effective for processing high-dimensional or sequential data, albeit with substantial computational requirements \cite{jieyang2023systematic}.

Convolutional Neural Networks excel in tasks like image recognition due to their automatic feature detection capabilities, although they require considerable computational resources \cite{ribeiro2022fault}. Recurrent Neural Networks are particularly well-suited for time-series data because of their capacity to store sequential information; however, they are prone to the vanishing gradient problem, which affects prediction accuracy over long sequences \cite{isermann2006fault}. \emph{Hidden Markov Models} (HMMs) are also effective for capturing sequential dependencies, although they are unsupervised and necessitate large datasets for training \cite{isermann1993fault, frank1997survey}.

Signal sampling and acquisition are foundational in digital technology, with applications from control systems to digital communication. Traditional sampling employs a uniform time-based cycle, known as \emph{Riemann Sampling} (RS), which integrates along the time axis. This method is widely used for its simplicity in analysis and implementation \cite{yan_battery_2019, reyes2018just}. In contrast, \emph{Lebesgue Sampling} (LS) is an event-driven approach where data points are recorded only when a measurement transitions between states. This approach, which integrates along the y-axis (i.e., measurement value), reduces the number of data points and is ideal for systems with limited computational resources \cite{reyes2018just}.

In RUL prediction, physical quantities indicating equipment health gradually change over time due to wear or aging. In early operational phases, signals typically remain stable with minimal variation, making continuous monitoring through RS both redundant and resource-intensive. In such cases, Lebesgue-based sampling techniques offer a novel solution, especially when RUL prediction is constrained by computational limits \cite{reyes2018just}. By focusing on state changes rather than continuous time-based measurements, LS-based sampling in hybrid models can reduce data volume while maintaining predictive accuracy, making it well-suited for edge devices with limited processing capacity.

\emph{Hybrid models} combining Lebesgue Sampling with traditional data-driven or model-based methods benefit from event-driven data acquisition, allowing for efficient resource allocation without compromising RUL prediction quality. This sampling adaptation enhances the hybrid model's ability to manage large datasets more efficiently, particularly under constrained computational conditions, contributing significantly to predictive maintenance solutions.

\begin{table}[ht!]
    \centering
    \renewcommand{\arraystretch}{1.5}
    \begin{tabular}{|p{.25\textwidth}|p{.28\textwidth}|p{.35\textwidth}|}
        \hline
        \textbf{Algorithm} & \textbf{Advantages} & \textbf{Limitations} \\ \hline
        Logistic Regression & Simple, efficient & Assumes linearity \\ \hline
        Artificial Neural Network\newline (ANN) & Robust to noise, stores\newline patterns & Unpredictable behavior \\ \hline
        Support Vector Machine\newline (SVM) & Effective in high dimensions & Not suitable for large datasets \\ \hline
        Deep Belief Network\newline (DBN) & Quick training on small\newline datasets & Network structure complexity \\ \hline
        Convolutional Neural\newline Network (CNN) & Excels in feature detection & Computationally demanding \\ \hline
        Recurrent Neural Network\newline (RNN) & Effective for time-series data & Suffers from gradient vanishing \\ \hline
        Hidden Markov Model\newline (HMM) & Captures sequential\newline dependencies & Requires large datasets \\ \hline
        Cluster Visualization & No need for labeled data & Lacks standardised parameter tuning \\ \hline
    \end{tabular}
    \caption{Comparison of Predictive Maintenance Algorithms based on shallow and deep learning approaches \cite{hu2020battery, ge2021review, hasib2021comprehensive, jieyang2023systematic}.}
    \label{table:Comparison_PredictiveMaintenance}
\end{table}

As shown in Table~\ref{table:Comparison_PredictiveMaintenance}, each algorithm presents distinct strengths and limitations, and their selection depends on the system's specific data and operational requirements. These predictive maintenance approaches, particularly with the integration of Lebesgue Sampling in hybrid models, contribute significantly to enabling real-time, data-driven maintenance solutions across various industrial applications.

\subsection{System design}
\label{subsec:Architectur}
The chapter discusses fault-tolerant system design in safety-critical and highly automated environments. It provides an overview of key strategies and architectural principles needed to ensure system reliability, availability and safety under failure conditions. Section~\ref{subsec:Concepts} introduces core fault tolerance concepts, including \emph{fail-safe}, \emph{fail-degradation}, and \emph{fail-operational strategies}. Section~\ref{subsec:Redundancy} explores redundancy techniques and their impact on system reliability and availability. Section~\ref{subsubsec:System Architectur} outlines different system architectures, highlighting \emph{domain} and \emph{zonal approaches}. Finally, Section~\ref{subsec:rekonf} focuses on \emph{dynamic reconfiguration}, emphasizing real-time adaptation to maintain system functionality.

\subsubsection{Fail-safe, fail-degradation and fail-operational concepts} \label{subsec:Concepts}

\begin{figure}[!ht] 
    \centering                  
    \def\svgwidth{.8\textwidth}    
    \input{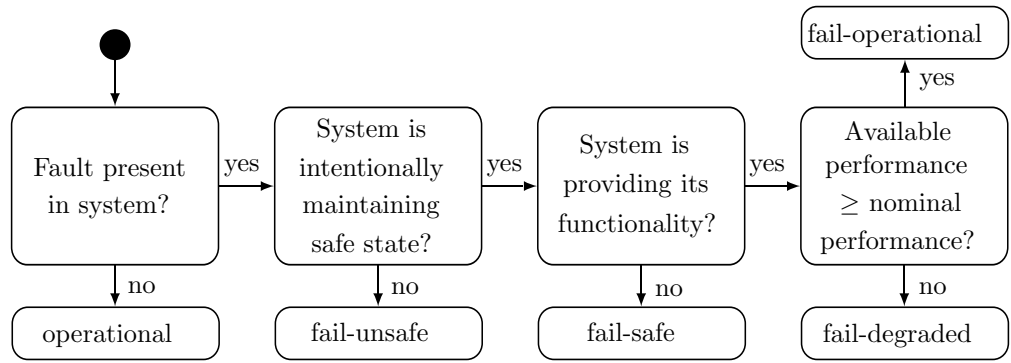} 
    \caption{Comparison of Fail-Safe, Fail-Operational, and Fail-Degradation strategies~\cite{Stolte.2022}.} 
    \label{fig:failestates}          
\end{figure}

Fault tolerance is a critical aspect of modern system design, especially in safety-critical and highly automated environments~\cite{Kohn.2015, Ishigooka.2018, Schmid.2019, Sari.2020}. As systems become more complex, integrating numerous electronic components and software subsystems, the likelihood of failures increases. Faults in one component can cascade through interconnected systems, leading to catastrophic consequences if not properly managed. For example, in autonomous vehicles, the failure of a single sensor or control module could potentially lead to dangerous driving conditions if not mitigated. Therefore, fault tolerance ensures that the system can continue to operate safely and effectively, even when individual components fail. To achieve fault tolerance, three key concepts: Fail-safe, fail-operational, and fail-degradation provide distinct approaches to managing failures and ensuring system reliability and availability~\cite{Stolte.2022}. To further clarify these strategies, Fig.~\ref{fig:failestates} is illustrating how fail-safe, fail-operational, and fail-degradation approaches differ from each other and how they work together to create a fault-tolerant system.

Fail-Safe systems transition into a predefined safe state when a failure occurs~\cite{vomDorff.2020}. This safe state is designed to minimize risks and prevent any further harm. In automotive systems, this often means ceasing functionality to avoid dangerous situations, such as in braking systems where mechanical fallback mechanisms may be activated upon failure of electronic components. Although the system stops functioning, fail-safe strategies ensure that safety is prioritised over continued operation. The focus is on bringing the system to a state that poses no unreasonable risk, even if this means halting operations entirely.

Fail-Operational systems maintain their full functionality even after a fault has occurred~\cite{Weiss.2020}. This is crucial in applications where uninterrupted operation is essential, such as in autonomous driving or aerospace systems. Fail-operational systems are designed to continue providing their specified functionality without performance degradation. For instance, in autonomous vehicles, even if one sensor fails, the vehicle must continue to operate using redundant sensors to complete its mission or reach a safe state. Fail-operational strategies ensure that critical systems can continue functioning without a loss in performance, enhancing both system availability and safety.

Fail-Degradation provides a middle ground, allowing systems to continue operating with reduced performance in the event of a failure~\cite{Yu.2018}. Unlike fail-operational systems, which maintain full functionality, fail-degradation allows for impaired functionality while still ensuring safety. In the automotive context, a fail-degradation strategy could involve reducing the speed of an autonomous vehicle if its primary navigation system fails, allowing it to safely continue driving at a reduced capability or move to a safer location. This concept is especially useful when full operational redundancy is too expensive or unnecessary. Fail-degradation increases both reliability and availability by ensuring that the system can still function, albeit at a lower capacity, instead of shutting down entirely. Fail-degradation can also be part of fail operational concepts~\cite{Weiss.2020}.

\subsubsection{Redundancy and its role in fault tolerant systems} \label{subsec:Redundancy}

Redundancy is fundamental to achieve fault tolerant capabilities in safety-critical systems~\cite{J..2020}. By introducing multiple components or subsystems that can take over in case of failure, redundancy ensures that the system remains operational or enters a safe state when faults occur. There are different strategies for implementing redundancy, often through \textbf{m-out-of-n} systems, where $n$ components are in operation, and the system continues functioning as long as at least $m$ components are working~\cite{Ivanova.2022, vanGemund.2012}. 

In redundant systems, \textbf{NooN} represents a \textit{serial configuration}, where all \(N\) components must function correctly for the system to operate. This configuration offers high reliability but is vulnerable, as any single failure causes the entire system to fail. On the other hand, \textbf{1ooN} represents a \textit{parallel configuration}, where the system can function as long as at least one of the \(N\) components works.

\textbf{Reliability} focuses on the probability that the system will operate without failure over a specified time. For a redundant system with \( n \) identical components, each with reliability \( R_i(t) \), the overall system reliability \( R_s(t) \) for an \textit{m-out-of-n} configuration is given by:
\begin{equation}
R_s(t) = \sum_{k=m}^{n} \binom{n}{k} R_i^k(t) \left( 1 - R_i(t) \right)^{n-k}
\end{equation}
This formula calculates the probability that at least \( m \) components are functioning correctly, ensuring system operation. Redundancy in \textit{fail-safe} systems guarantees that, upon failure of one or more components, the system transitions into a \textit{safe state}. The goal in such systems is not to continue operation, but rather to prevent unsafe conditions by shutting down or moving to a controlled, safe condition.

\textbf{Availability}, on the other hand, is crucial for \textit{fail-operational} systems, where the system must remain functional even after failures. Availability is a measure of the system’s ability to perform its intended function at any given time, factoring in maintenance and repair periods. In \textit{fail-operational} systems, maintaining availability is often more critical than ensuring perfect reliability, as the system needs to continue functioning despite component failures. The \textit{availability} of a redundant system is influenced by both its reliability and its ability to be repaired or reconfigured, and is asymptotically calculated as:
\begin{equation}
A_s = \frac{MTTF}{MTTF + MTTR},
\end{equation}
where \( MTTF \) is the \textit{mean time to failure}, and \( MTTR \) is the \textit{mean time to repair}. In this context, redundancy allows the system to continue operating while repairs are made, ensuring minimal downtime. The effectiveness of repair or reconfiguration in fault-tolerant systems is directly influenced by the ability to separate or diversify system elements~\cite{Safari.2022, Stetter.2020}. 

\textit{Separation} refers to the physical isolation of redundant subsystems, ensuring that a failure in one part of the system does not affect others. For example, in aircraft systems, placing critical sensors or control units in different parts of the vehicle prevents damage in one area, such as from fire or impact, from compromising the entire system. This isolation allows faulty components to be repaired or replaced while the rest of the system remains operational, thus maintaining availability~\cite{Safari.2022, Stetter.2020}.

\textit{Diversification}, on the other hand, involves using different technologies, designs, or suppliers for redundant components to reduce the risk of common-cause failures. In a fail-operational system, diversity ensures that if one component fails due to a specific failure mode or environmental condition, another component with a different design or tolerance can continue functioning. For instance, in autonomous vehicles, using a mix of sensors like lidar, radar, and cameras ensures that the system can maintain functionality even if one type of sensor is compromised by conditions like fog or rain~\cite{Safari.2022, Stetter.2020}.

In \textit{fail-degradation} systems, redundancy enables the system to degrade gracefully in response to failures, continuing to function at a reduced capacity rather than completely ceasing operation~\cite{Stolte.2022}. For instance, if several sensors fail in an autonomous vehicle, the remaining sensors may allow the vehicle to continue driving, although with reduced accuracy or slower speeds. This kind of redundancy ensures that the system can maintain essential functions and prevent complete shutdown, balancing \textit{reliability} and \textit{availability}.

\subsubsection{System architectures} \label{subsubsec:System Architectur}

The concepts of redundancy, separation, and diversification are foundational to designing robust electronic control units (ECUs), particularly in safety-critical systems such as automotive and aerospace applications~\cite{Stetter.2020}. In practice, ECUs use redundancy strategies to maintain operation in the face of failures. For instance, a 1ooN configuration may be employed for non-critical systems where basic functionality must continue despite component failures, while a 2oo2 or 2oo3 configuration could be used in safety-critical systems that require more stringent fault-tolerance measures~\cite{Kohn.2015}. By deploying multiple ECUs in parallel, the system can isolate faults or switch to a backup ECU if necessary~\cite{Krishnamoorthy.2023}. Two different fail-operational ECU architectures are shown in Fig.~\ref{fig:ECU} in Appendix~1.



\begin{figure}[!ht]
    \centering
    \begin{subfigure}[b]{0.48\textwidth}
        \centering
        \def\svgwidth{\textwidth}    
        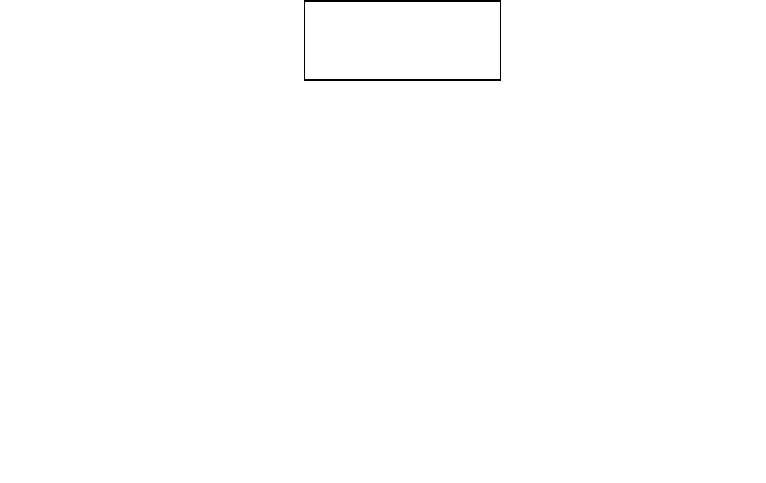 
        \caption{Domain Architecture}
        \label{fig:image3}
    \end{subfigure}
    \hfill
    \begin{subfigure}[b]{0.48\textwidth}
        \centering
        \def\svgwidth{\textwidth}    
        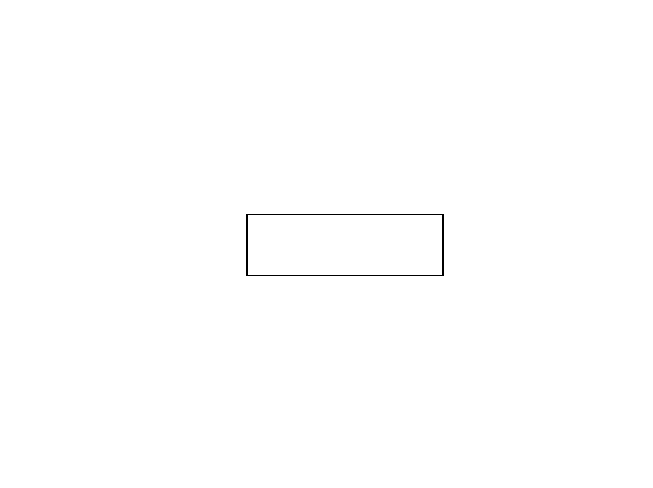 
        \caption{Zonal Architecture}
        \label{fig:image4}
    \end{subfigure}
    \caption{Electronic Control Unit Architectures at the System Level}
    \label{fig:S-Arch}
\end{figure}

ECU redundancy schemes can be integrated into broader architectures at the system level, which are commonly organised into either domain architectures or zonal architectures~\cite{Krishnamoorthy.2023, Rumez.2020}. These system architectures dictate how ECUs are distributed, communicate, and handle redundancy across the entire System. The two approaches are illustrated in Fig.~\ref{fig:S-Arch}.

In a \textit{domain architecture}, ECUs are organised based on their functional roles, with each domain representing a specific set of functions. For example vehicle, the powertrain domain manages functions related to the engine, transmission and drivetrain. The body control domain oversees lighting, doors, windows, and climate control. The infotainment domain handles multimedia, navigation, and user interfaces and the safety domain manages systems such as airbags, ABS (anti-lock braking system), and electronic stability control. In this architecture, each domain operates somewhat independently, with ECUs dedicated to specific tasks within that domain. Redundancy strategies like 2oo3 TMR or 2oo2 DFS can be applied within each domain to ensure critical functions like powertrain or safety can continue to operate even in the event of a failure. The advantage of a domain architecture is that it allows for easier management and development, as each domain is functionally isolated, allowing for specialization and optimization of ECUs within that domain. However, a significant drawback is that communication between different domains, such as powertrain and safety, requires more complex wiring and can introduce latency. Physical separation between ECUs in different domains may also limit fault tolerance across domains, as sharing resources in failure conditions is more difficult.

A \textit{zonal architecture}, on the other hand, organizes ECUs based on their physical location in the vehicle, dividing the system into zones. For instance, the front zone of a vehicle would include ECUs handling headlights, front sensors, and climate control for the front passengers, while the rear zone would include ECUs for tail lights, rearview cameras, and rear passenger controls. Each zone houses ECUs that manage multiple functions from different domains. For example, an ECU in the front zone could handle both body control functions like headlights and safety functions like front sensors. In this approach, redundancy strategies like 2oo2 DFS or 2oo3 TMR can be applied within or across zones to improve fault tolerance and ensure continuous operation. One of the key advantages of a zonal architecture is that it reduces wiring complexity, as ECUs in the same zone can communicate and share resources locally, improving efficiency. Real-time communication between different systems within a zone is also faster, as there is less distance and latency in the connections. Moreover, zonal architectures allow for better scalability, as new zones or systems can be added without redesigning the entire vehicle architecture. However, this approach makes it more challenging to specialize ECUs for specific functions, since each ECU within a zone may need to handle multiple tasks across different domains, increasing the complexity of the system. Additionally, ensuring redundancy and fault tolerance for critical systems that span across zones can introduce challenges in managing inter-zone communication and resource sharing.

\subsubsection{Dynamic reconfiguration in automated systems} \label{subsec:rekonf}

As systems become increasingly automated, dynamic reconfiguration—the ability to adapt the configuration of electronic control units (ECUs) in real time—plays a critical role in ensuring continuous operation~\cite{Stoll.2021, oszwald2018dynamic}. This capability is particularly important in fail-operational systems, where safety and availability are paramount. Dynamic reconfiguration allows the system to respond to faults or failures by redistributing tasks or switching over to backup resources. 

In highly automated systems, where there is no human fallback option, dynamic reconfiguration becomes crucial. As, e.g. autonomous vehicles, takes full control over critical functions, the system must have the ability to adapt in real time to any failures or faults~\cite{Horcas.2018}. This is especially important in fail-operational systems, where the vehicle must continue functioning even after a component failure, without relying on human intervention. The shift towards fail-operational systems requires transitioning from static E/E-architectures to more dynamic solutions that can handle real-time reconfiguration~\cite{oszwald2018dynamic}. Current Electric/Electronic (E/E) architectures typically implement a fail-safe mode, where the driver must intervene in case of system failure. However, as the industry moves towards higher levels of automation (SAE levels 4 and 5), necessitating the development of fail-operational architectures that can dynamically reconfigure to ensure the system remains operational without human intervention. Dynamic reconfiguration is realised through both hardware and software adaptations, allowing ECUs to switch tasks or functions between redundant components. For example, in a fail-operational system for brake and steering control, multiple ECUs work together, and if one ECU fails, another ECU—previously in a hot standby mode—takes over the failed ECU's tasks~\cite{oszwald2018dynamic}. 

Dynamic reconfiguration can be effectively modeled using \textbf{Markov processes}, which are suited to describe probabilistic transitions between system states based on failures, reconfigurations, or repairs~\cite{Djoudi.2022, Haring.2022}.  An example of a Markov transition diagram of a 2-out-of-3 System with failure rates $\lambda_i$ and repair rates $\mu_i$ can be found in Appendix~2, Fig.~\ref{fig:markov}.

In such a system, the state of each ECU evolves based on transition probabilities that account for different operational conditions. The probability of transitioning from one state \(i\) (e.g., fully operational) to another state \(j\) (e.g., reconfiguration) at a time \(t\) is given by:
\begin{equation}
p_{ij}(t) = P(\text{transition from state } i \text{ to state } j \text{ at time } t)
\end{equation}
This transition probability captures how likely the system is to reconfigure itself in response to detected failures.

The \textbf{reconfiguration time} is a critical factor, defined as the time it takes to detect a fault and switch control to the standby ECU. According to~\cite{oszwald2018dynamic}, the \textbf{fault reaction time} (FRT), which includes both fault detection and reconfiguration time, must be shorter than the \textbf{failure tolerance time interval} (FTTI) for the system to maintain safe operation. For example, in a braking system, if the FTTI is 160~ms, the reconfiguration process must occur within that time to prevent hazardous events, as outlined by ISO 26262 standards.

To further quantify this, the system's \textbf{transition rate matrix} \(Q\) describes the rate of transitioning between states, where \(q_{ij}\) represents the rate at which the system moves from state \(i\) to state \(j\). For instance, the transition from normal operation to a reconfigured state (following ECU failure) can be described as:
\begin{equation}
q_{ij} = \lambda_{ij} e^{-\lambda_{ij} t}
\end{equation}
where \( \lambda_{ij} \) is the rate of failure or reconfiguration. The system must then balance between fail-operational and fail-safe states based on real-time conditions.

In practice, dynamic reconfiguration ensures that critical systems like braking and steering remain operational, even in highly automated vehicles. \cite{oszwald2018dynamic} describes a three-ECU system for braking and steering, where the third ECU serves as a backup and is dynamically activated in the event of a failure in one of the primary ECUs. This approach provides both redundancy and flexibility, allowing the system to reconfigure in real time while adhering to strict timing requirements, as outlined in the failure detection Time and reconfiguration time tables provided in~\cite{oszwald2018dynamic}. For example, if a nominal steering function is lost, the system must detect the failure within 30~ms and reconfigure within 100~ms to maintain safety and operational integrity.

By leveraging real-time monitoring, predictive maintenance, and Markov-based probabilistic modeling, the system can anticipate and mitigate failures before they escalate into critical issues. This aligns with the fail-operational concepts we have previously discussed, ensuring that the system continues to function even when individual components experience faults.

\subsection{Simulation-based reliability analysis}
\label{subsec:Simulation}

In reliability engineering, simulation techniques are critical tools for modelling, analysing, and improving system reliability \cite{naikan2016review,ram2019modeling}. 
They allow engineers to predict the performance and failure behaviour of complex systems under different operating conditions, without the need for time-consuming or expensive physical testing. 
Simulation analysis is particularly relevant for safety-critical systems involving humans, where experimentation and testing is not possible for safety, ethical or cost reasons. 
A typical example is automated driving. 
In fact, the probability of a traffic accident per kilometre driven is extremely low, while testing and experimenting new systems in real traffic situations is difficult and expensive \cite{kalra2016driving}. 
Generally speaking, simulation-aided analysis can help optimise maintenance schedules, component designs, and system architectures for maximum reliability or test different scenarios in a \emph{what-if} analysis.
Simulation techniques are especially relevant for analysing the reliability of dynamic mechanical systems that exhibit wear and fatigue effects, or more generally systems where the failure rates are time-dependent \cite{naikan2016review,ram2019modeling,dong2020structural,feyzi2021review}. 
Indeed, such systems cannot be modelled using Markovian processes and are in general not analytically tractable. 
Simulations can also handle the complexity of real world systems with multiple interacting components, for instance by identifying critical components, where redesigns can enhance reliability \cite{maurya2020reliability,friederich2024reliability}.
Simulation techniques are also very useful for scheduling of complex maintenance strategies for repairable systems \cite{de2020review,van2022predictive,errandonea2020digital}. 
They can also incorporate sensor data and be used as digital twins for real-time monitoring of the system health \cite{hu2021digital,liu2021review,bado2022digital,liu2022digital,bofill2023exploring}.

Simulation techniques in reliability engineering showing particular innovations during the last decades include
\begin{itemize}
    \item {\bf Monte Carlo Simulation} (MCS), one of the most widely used simulation techniques in reliability engineering (see the reviews \cite{qiu2013structural,afshari2022machine,song2023monte}). 
    Monte Carlo techniques are particularly well suited to discrete-event simulation.
    MCSs use random sampling to simulate the behaviour (typically failure times) of complex systems, especially when analytical solutions are difficult or impossible. 
    For example, they can address the structural reliability of multi-component systems including degradation processes \cite{lin2015reliability,lyu2023reliability} or with interdependent failure rates \cite{bian2021reliability,julitz2024reliability}. 
    
    \item {\bf Fault Tree Analysis} (FTA) Simulation, a top-down, deductive failure analysis tool used to identify potential causes of system-level failures. 
    FTA methods can be used to quantify the likelihood of different failure scenarios and to identify critical components \cite{shi2015fault}.
    FTA is particularly useful in safety-critical systems, such as aerospace or nuclear power, where understanding the root cause of potential failures is important \cite{abdelghany2021event,maurya2020reliability,mamdikar2022dynamic}, see also the bibliographic review \cite{yazdi2023fault}.
    
    \item {\bf Finite Element Method} (FEM), a numerical technique widely used in reliability engineering to analyse and predict the behaviour of physical systems described by (nonlinear) partial differential equations.
    FEM is especially relevant when dealing with complex structures, materials, and loading conditions \cite{feyzi2021review,ereiz2022review,kudela2022recent}. 
    FEM allows engineers through numerical schemes to break down complex systems into smaller, simpler parts (finite elements) to study stress, strain, heat distribution, vibration, and other physical phenomena that can affect the reliability of a system.
    
    \item {\bf Multi-Agent Systems} (MAS), to analyse the reliability of complex systems where multiple autonomous components (or subsystems) must work together to maintain system reliability. 
    MAS can dynamically adjust to changes, detect failures, and coordinate repairs or optimisations in a distributed and autonomous manner. 
    They are generally based on a reinforcement learning processes for which model parameters are adjusted through feedback reward mechanisms \cite{zhang2021multi}.
    Typical applications concern robotics and automated (smart) manufacturing \cite{o2022improving,su2022deep,nazabadi2024joint}. 
    MAS are especially suited for modelling emergent collective behaviours such as cascading  failures \cite{liu2022risk,xu2022failure}.
    
\end{itemize}

\subsubsection{Digital twin}
\label{subsec:DigitTwin}

In reliability engineering, a digital twin is a dynamic, virtual model of a physical system that uses real-time data, simulation, and analytics to predict system performance and behaviour. The digital twin concept is becoming increasingly important as it enables more accurate, detailed, and real-time analysis of system reliability, including failure prediction, maintenance scheduling, and optimisation of system performance, see the reviews \cite{bado2022digital,thelen2022comprehensive,thelen2023comprehensive}.
A digital twin is created using data from sensors, historical information, and models of the physical system, often differential models. 
This virtual model is continuously updated with real-time data from the physical counterpart, allowing engineers to monitor, analyse, and optimise the physical system in real time.
In the context of reliability engineering, digital twins are particularly useful for:
\begin{itemize}
    \item Monitoring system health and performance in real time \cite{liu2022digital},
    \item Predicting failures using predictive analytics \cite{liu2021review,bofill2023exploring},
    \item Optimising maintenance strategies (e.g., predictive maintenance) \cite{errandonea2020digital,van2022predictive},
    \item Improving reliability by simulating and testing different configurations, conditions, scenarios or repair strategies \cite{rosebrock2020simulation,hu2021digital,bado2022digital}.
\end{itemize}
Applications of digital twins in reliability engineering cover many areas, including aerospace and aviation, manufacturing, energy and power systems, automotive industry, smart cities and infrastructure, etc., see the reviews \cite{botin2022digital,liu2021review,hu2021digital}.
Typical examples are in automotive engineering for accident analysis and reconstruction of traffic accidents. The modelling of the vehicle structure and the deformation due to impacts and collisions can be performed at low cost using continuous physical models and FEM simulation analysis \cite{evtiukov2018finite,noorsumar2022mathematical}. 
Another example of applications is the field of automated vehicles and advanced driver assistance systems. 
Here, the extremely low probability of accidents in traffic and the obvious safety concerns prevent testing and experimentation under real conditions and are generally carried out using numerical simulation, see, e.g., the review \cite{dona2022virtual}.

\subsubsection{Accelerated life testing}
\label{subsec:AccLifeTest}
For safety-critical technical products or systems, the term \emph{reliability} is defined as a characteristic or behavioural trait that indicates the extent to which a specific function is performed in a reliable manner within a given time interval \cite{bracke2024reliability}. To determine the exact lifespan for purposes such as maintenance intervals, service policy or optimal operating conditions, it is essential to conduct targeted tests that accurately reflect actual failure behavior. Nevertheless, given that the anticipated product lifespan can span several years, accelerated life testing is employed to facilitate the assessment of the planned service life within a shorter time frame. 

As part of the Design of Experiments (DoE), targeted stresses and loads are simulated by adjusting individual or multiple parameters in combination, thereby accelerating the ageing process or degradation. The challenge in simulating degradation or failure behavior lies in replicating or representing the causative factors of damage responsible in reality at an earlier point in time, despite the accelerating factors \cite{WayneNelson.1990,Siebertz.2017}. A basic distinction is made here between qualitative High Accelerated Testing (HALT) and quantitative Accelerated Testing (ALT). While HALT aims to provoke failure causes as quickly as possible through overloading, ALT is conducted strictly within specified product specification limits, making it transferable to real-world conditions. Although sensor data can be collected during HALT, it only allows for qualitative assessment, whereas ALT can be used for quantitative modelling. 

While \cite{Hinz.2014} categorises the possible forms of accelerated life testing in more detail, \cite{WayneNelson.1990} and \cite{bracke2024reliability} provides a methodological overview of the quantitative modelling types depending on the degradation or damage event. Specific applications include, for example, modelling the complex degradation behaviour of corroding lead anodes (see reference \cite{ChristophRosebrock.2023}), or deriving an acceleration factor (scuffing factor) relevant for the maintenance forecast of products from the construction industry that are designed to last longer than 19 years \cite{FabianHartwig.2023}. 

\subsubsection{Scenario-based simulation}
\label{subsec:ScenarioBased}

Scenario-based simulation is a widely used approach for evaluating the reliability of complex systems, such as those encountered in safety-critical domains \cite{naikan2016review,tang2023survey}. 
These systems are typically characterised by numerous interacting components and dynamic behaviour, leading to high-dimensional problems that are often NP-hard. The complexity of such systems can escalate rapidly, making it infeasible to derive general solutions analytically. Consequently, it is essential to constrain the analysis space and focus on identifying specific safety-critical scenarios. This process requires domain expertise to determine which scenarios pose significant risks to the system. For example, in the context of autonomous vehicles, understanding critical traffic situations such as lane changes or interactions at intersections is fundamental for ensuring system safety. Identifying these scenarios enables a targeted evaluation, facilitating more effective safety assessments \cite{sun2021scenario,tang2023survey}.

In \cite{Grassler.2020}, the authors categorise scenario methods into three main groups. {\bf Intuitive Logics} is a qualitative approach, introduced by~\cite{Kahn.1967}, relies on expert judgment to develop scenarios without using mathematical models. It focuses on generating diverse future environments to explore strategic alternatives and assess potential impacts, making it suitable for exploratory long-term planning. {\bf Cross-Impact Analysis} uses probabilistic modeling to analyze interdependencies between influencing factors. Starting from initial probabilities for each factor, it computes how changes in one factor affect the probabilities of others using a cross-impact matrix. Scenarios are then generated based on the highest combined probabilities, making it a mathematically rigorous approach for identifying the most likely future developments. {\bf Consistency-Based Approaches} evaluate the pairwise consistency between different projections to create scenarios that are logically coherent. By scoring how well different assumptions fit together, it identifies consistent sets of scenarios, ensuring internal alignment and plausibility, but without emphasizing likelihood of occurrence.

\section{Challenges and development perspectives}
\label{sec:Challenges}

Although sensor-based monitoring, system design and simulation-based reliability analysis have provided important advances and innovations in reliability engineering over the last decades, many challenges remain to be addressed. 
In this section we suggest some of the more important ones.

\subsection{Trustworthy AI and real-time data processing}
\label{subsec:SafeAI_BigData}

Ensuring that machine learning systems perform as expected, minimising risks and operating reliably under different conditions is crucial in advanced reliability methods. 
This aspect is commonly referred to as \emph{Trustworthy} or \emph{Safeguarding AI} in engineering \cite{kaur2022trustworthy,chamola2023review,ramezani2023scalability}.
With the increasing integration of machine learning algorithms into safety-critical systems (e.g., transportation or healthcare) and  industrial automation, failures can lead to catastrophic outcomes. 
Therefore, it is necessary to quantify confidence and uncertainty in AI predictions and to explore the robustness of the algorithms in different situations, and especially rare ones.
Safeguarding AI in reliability engineering includes various strategies and frameworks to ensure safe and consistent performance \cite{thelen2023comprehensive}. 
The following are some recently developed methods and concepts.

\begin{itemize}
\item{\bf Robust and Redundant AI}
~Key considerations for safeguarding AI are robustness, ensuring that AI systems can handle unexpected inputs or disturbances without failing, and redundancy, when incorporating multiple AI algorithms or fallback mechanisms to ensure that if one system fails, another can take over. 
These aspects remain challenging. 
They can be addressed through Monte Carlo simulation, fuzzy logic and ensemble learning methods \cite{mahmood2013fuzzy,song2023monte}, and in system design through redundant, fail-safe, fail-operational or fault-tolerant architectures that drive AI systems to safe or operational states in case of failure.

\item{\bf Explainable and Transparent AI}
~The explainability and transparency are also key considerations, by building AI models that are interpretable so that engineers can understand why a certain decision was made, and by ensuring that the development, training, and operation of AI systems are documented \cite{ramezani2023scalability}. 
This is especially important in safety-critical applications. 
Indeed, as AI models become more complex, particularly with the use of ensemble learning or deep learning, their decisions become harder to explain and trust in safety-critical applications.
A typical application concern the driving automation where AI algorithms for vehicle perception and motion planning quickly turn out to be black boxes whose operation is difficult to interpret.
More research is needed into explainable AI (XAI) techniques to ensure that human operators can understand and trust AI-based systems in high-stakes environments \cite{arrieta2020explainable,narteni2021explainable,chamola2023review}.

\item{\bf AI Model Validation and Verification}
~Model validation is an important task of AI modules, by ensuring that the AI model performs well under various operational conditions and uncertainties \cite{jiang2007bayesian}. 
Model verification, by checking that the AI system adheres to its specified requirements and does not deviate from expected behavior, even in dynamic or unpredictable environments, is also fundamental for trustworthy AI \cite{dona2022virtual,li2022features}.

\item{\bf Handling of Rare Events}
~Another challenge is the handling of rare events. Indeed, AI systems often struggle with rare or outlier events, especially in reliability engineering where catastrophic failures are rare but highly impactful \cite{afshari2022machine,shyalika2023comprehensive}.
Developing models that can reliably predict rare events is still a challenge. 
Methods like extreme value theory, enhanced Monte Carlo simulations, or specialised outlier detection algorithms are needed to better handle these edge cases.

\item{\bf Scalability and real-time processing of AI Models}
~The scalability of AI models is also especially challenging. In fact, many AI models, such as networks or ensemble methods, are difficult to scale for large, complex systems with many interacting components \cite{zio2016some,belgaum2021role,ramezani2023scalability}.
There is an important need for more efficient algorithms and computational techniques to handle the large-scale, high-dimensional systems found in industries like aerospace, healthcare, and energy.
Another challenge is to make AI model processing operational in real time. 
This challenge is particularly relevant for smart manufacturing, autonomous vehicles and other safety-critical dynamic systems.
\end{itemize}



\subsection{Addressing system complexity}
\label{subsec:SystComplex}

Dealing with system complexity is an important task in reliability engineering. 
In fact, the operation of complex systems often consists of combination of different modules and components interacting with each other. 
The system architecture may also be complex and include state and time dependent mechanisms, for instance for redundant, safe tolerant or fail safe designs. 
To address the reliability of these systems, the complexity of the system must be assessed, taking into account the system architecture. 
This task is not straightforward, especially when the system modules and components are interconnected. 
The interdependence of the component failures, such as \emph{common cause failures} and \emph{cascading failures} makes the reliability analysis much more complex. 
Furthermore, the failure interdependency can drastically affect the system reliability \cite{kotz2003effect}, for example for the perception or motion planning of autonomous vehicles \cite{gottschalk2022does,icaart21,julitz2024reliability}.
Indeed, dependent components require the system to be considered as a whole (i.e., in a multidimensional framework), since the components cannot be regarded individually.
Several models have been developed for this purpose, including lifetime distribution models \cite{zhang2017component,navarro2021redundancy,kelkinnama2022optimal},  
system state models \cite{mi2018reliability,li2023redundancy},  
failure interaction models \cite{cai2012using,fang2016allocating},
failure propagation models \cite{da2007active}, see also the review \cite{zeng2023dependent} and references therein.
However, the task remains challenging and is and currently under active development \cite{fang2016allocating,julitz2024reliability}. 

Another challenge in addressing system complexity relates to the diverse technologies involved in system reliability, particularly for cyber-physical systems. 
The combination of IoT, AI, robotics and digital control can lead to unforeseen failure modes that traditional reliability methods may not be able to address. 
Another challenging component of complexity is the collaboration between humans and machines in sociotechnical systems. 
This aspect is explored below.

\subsection{Sociotechnical systems}
\label{subsec:HuamFactor}

A sociotechnical system can be defined as a system consisting of two interacting components: a technical component and a social component (see, e.g., \cite{ulich2001arbeitspsychologie}). The concept emphasises that both components of an overall system have to be addressed when changes are introduced in one part – due to the complex interaction structure (see, e.g., \cite{walker2008review}). Two aspects are mainly relevant in sociotechnical theory: 
\begin{enumerate}
    \item Interactions between technical and social subsystems determine the overall system success and
    \item Optimising either social or technical component without considering the other part or the interaction can lead to unintended, non-linear relationships diminishing overall system performance \cite{walker2008review}.
\end{enumerate} 
The ideas stem from original coal mining workplace studies, where technological innovations meant to improve working situations and increase performance, but introduced unintended side-effects of changing social structures among the workers \cite{trist1951some}. Outcomes were high fluctuation and sick-leaves instead of performance increases \cite{liggieri2019mensch}. The theory has implications for any change in case of complex systems. Taking a look at the changes of industry 4.0, several aspects related to sociotechnical theory have to be taken into account.

First of all, the increasing digitalisation and interconnectedness through e.g. cyber-physical systems increases the amount of available and relevant information and secondly autonomous and intelligent processes operating in real time, as well as interactions with virtual reality displays create new strain on workers \cite{frost2018neue}. Automation for example, poses the challenge of changing the role of the workers from active controllers to monitors which in turn leads to degradation in skills that might be necessary when abnormal situations occur \cite{bainbridge1983ironies}. This –along with the notion, that humans are not well suited for prolonged vigilance tasks– has already been formulated in the \emph{ironies of automation} \cite{bainbridge1983ironies}. The idea of compensating this challenge through training and getting the human to adapt to the new system challenges will not produce satisfactory outcomes. Instead, the sociotechnical systems approach shall be applied: before introducing technological innovations simultaneous estimation of social changes and influences on the human and possible countermeasures shall be addressed.






\section{Statistical learning methods}
\label{sec:Stat}

In this section we review statistical, learning and simulation methods for advanced reliability techniques of complex systems discussed in Sections.~\ref{sec:Innov} and \ref{sec:Challenges}.
The methods include unsupervised techniques for data visualisation and clustering in section~\ref{subsec:Unsupervised}. 
They also cover supervised techniques for prognostics, such as classification, regression and machine learning algorithms in section~\ref{subsec:Supervised}. 
See also \cite{bracke2024reliability} for in-depth reliability engineering basics. 

\subsection{Unsupervised learning methods}
\label{subsec:Unsupervised}

We first consider a multivariate dataset $X$ consisting of $n\in\{1,2\dots\}$ observations of $p\in\{1,2\dots\}$ variables (features)
\begin{equation}
X=\begin{bmatrix} x_1^1&x_1^2&\ldots&x_1^p\\[1mm]
x_2^1&x_2^2&\ldots&x_2^p\\
\vdots&\vdots&&\vdots\\
x_n^1&x_n^2&\ldots&x_n^p
\end{bmatrix}\in\mathbb R^{n\times p},\qquad \left|\begin{array}{ll}
      x_i=(x_i^1,\ldots,x_i^p)\in\mathbb R^p,&i=1,\ldots,n\\[1mm]
      x^j=(x_1^j,\ldots,x_n^j)^\top\in\mathbb R^n,~~& j=1,\ldots,p
      \end{array}\right.
\end{equation}
We assume that the variables $(x^1,\ldots,x^p)$ are dependent, i.e., the multivariate sample $X$ has a structure that we aim to explore.  
In general, the data are sensor measurements of various physical quantities such as temperature, pressure, vibration, speed (e.g. rotation), noise, etc.
The values describe the state of the system, e.g., a machine.

\subsubsection{Principal component analysis}
\label{subsubsec:PCA}
There exist many different graphical methods to visualise multivariate datasets, including parallel plots, radar charts, Andrews' plot \cite{andrews1972plots} or again Chernoff's face \cite{chernoff1973use}.
However, the dimension of the dataset, i.e. the number of features $p$ observed in the system, can be large, making it difficult to visualise the global structure of the data. 
In addition, several features may be highly correlated (e.g. noise with rotation speed) and the dimension of the dataset can be reduced without losing a significant amount of information. 

Principal component analysis (PCA) can be traced back to the pioneering work of Karl Pearson and, later, Harold Hotelling \cite{hotelling1933analysis}. 
The idea is to use a proper orthogonal decomposition of the data, the principal components, that maximises the amount of information.
In this way, the first principal component is the best representation of the data in one dimension in terms of variability, the first two components the best representation in two dimensions, and so on.
If the original multivariate dataset is highly correlated, the last components contain almost no information and can be neglected. 
Therefore, the dimensionality of the dataset can be reduced by using only the first components. 

\begin{figure}[!ht]
    \centering
    \includegraphics[width=.8\textwidth]{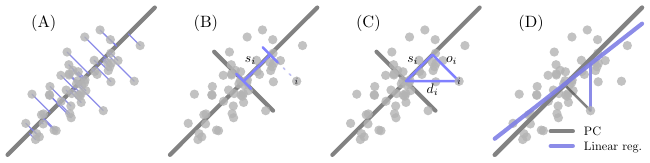}
    \caption{Principles of the principal components (example in two dimensions). (A): orthogonal projections of the data. (B): maximisation of the information on the axe $\sum_i s_i^2$. (C): As the total variability $\sum_i d_i^2=\sum_i s_i^2+\sum_i o_i^2$ is conserved, maximising $\sum_i s_i^2$ leads to minimising the squared orthogonal distances $\sum_i o_i^2$. (D): Principal component and linear regression generally do not match.}
    \label{fig:PCA0}
\end{figure}

Principal components are orthogonal linear combinations of the original variables. 
Geometrically, we maximise the variability of projections of the data (see Figure~\ref{fig:PCA0}). 
Thanks to the Pythagorean theorem, the total variability is conserved and maximising the variability on the axes resumes to minimising the sum of the squared orthogonal distances to the axes. 
This differs from the technique used in linear regression, which minimises the squared difference between a linear model and the data. 
For this reason, PCA is sometimes referred to as orthogonal regression.

Analytically, the principal components can be determined using basic linear algebra.
Let $\tilde X$ be the standard score transformation of the data with mean zero and variance one over the $p$ features and $\Gamma=\frac1n\tilde X^\top \tilde X\in\mathbb R^{p\times p}$ the variance/covariance matrix, i.e., $\Gamma_{j,k}=\text{covar}(\tilde x^j,\tilde x^k)$. 
Then, thanks to the diagonal formulation $\Gamma=P^{-1}DP$, where $D\in\mathbb R^{p\times p}$  isthe diagonal matrix of the ordered eigenvalues of $\Gamma$ and $P\in\mathbb R^{p\times p}$ is the matrix formed by the associated eigenvectors, the principal components are directly the columns of the matrix $\tilde XP\in\mathbb R^{n\times p}$. 
An example of application of PCA on measurements of banknotes is given in Appendix~3\footnote{The principal component analysis of banknote measurements can also be computed online at\\ \href{https://colab.research.google.com/drive/1D-Z-lWD_Z74Ts88EQfEjc2W1BLZBdqNw?usp=sharing}{\texttt{colab.research.google.com/drive/1D-Z-lWD\_Z74Ts88EQfEjc2W1BLZBdqNw}}}.

\subsubsection{Clustering techniques}
\label{subsubsec:Clustering}

Clustering techniques are unsupervised data mining methods designed to explore multivariate datasets and to identify patterns and clusters. 
In principle, the characteristics of the observations in a given cluster are more similar than those of the observations in the other clusters.
Clustering techniques can be distinguished according to the following criteria
\begin{itemize}
    \item Monitored and unmonitored clustering, whether the number of clusters is known or unknown
    \item Strict clustering, where each observation belongs to exactly one cluster
    \item Clustering with outliers, where some observations can also belong to no cluster 
    \item Overlapping clustering, where observations can belong to more than one cluster
    \item Fuzzy clustering, where each observation belongs to each cluster to some extent
    \item Hierarchical clustering, where observations of a child cluster also belong to the parent cluster
    \item Centroid clustering, where clusters are represented by a centroid (mean value)
\end{itemize}
Generally speaking, non-parametric clustering are empirical methods based on distance metrics between the observations or kernels. 
Parametric clustering rely on multivariate probabilistic distributions, typically multivariate normal distributions.
\medskip

\paragraph{\bf $K$-means clustering}
$K$-means algorithm in a non-parametric monitored strict clustering technique where a cluster is represented by a centroid. 
The terminology \emph{$K$-means} dates from the late 1960s \cite{macqueen1967some} although the method has been used since the 1950s.
Consider $(x_1,\dots,x_n)\in\mathbb R^{n,p}$ observations and a partition of clusters $S=\{S_1,\dots,S_K\}$ where $\mu_1,\dots,\mu_K$ are the means on each cluster. 
Then $K$-means clustering is the partition that minimises the inter-cluster variability:
\begin{equation}
    S_{\text{$K$-means}}=\text{arg}\min_S \sum_{k=1}^K\sum_{i\in S_k} \|x_i-\mu_k\|^2.
\end{equation}
In two dimensions, the $K$-means partition corresponds to a Voronoi diagram.
Note that since the total variability is constant for any clustering, minimising the inter-cluster variability is equivalent to maximising the intra-cluster variability.

\begin{figure}[!ht]
    \centering
    \includegraphics[width=.85\textwidth]{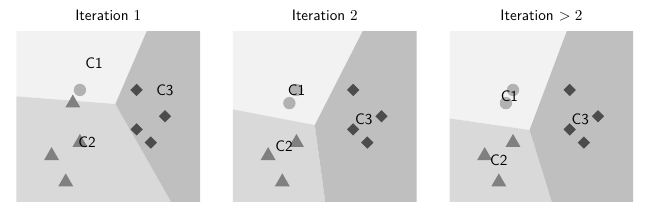}
    \caption{Iteration of the Lloyd-Forgy algorithm for the $K$-means clustering with $n=9$ synthetic bivariate observations. The algorithm converges in three iterations.}
    \label{fig:LF-algo}
\end{figure}

The $K$-means problem is computationally NP-hard difficult.
However, efficient heuristic algorithms quickly converge to a local optimum.
For instance, the {\bf Lloyd-Forgy algorithm} is an iterative process consisting of three steps:
\begin{enumerate}
\item initialisation: Choose $K$ random mean values $u_1^1,\dots ,u_K^1\in\mathbb R^p$.
\item Assignment: Each data object is assigned to the cluster with the smallest variance increment.
\begin{equation}
    S_k^t=\left\{x_i: \big \|x_i-u_k^t\big \|^2 \leq \big \|x_i-u_{k^*}^t\big \|^2\text{ for all }k^*=1,\dots ,K\right\},~ k=1,\dots,K.
\end{equation}
\item Update: Recalculate the centres of the clusters:
\begin{equation}
     u_k^{t+1}=\frac {1}{|S_k^{t}|}\sum _{i\in S_k^{t}} x_i,\qquad k=1,\dots,K.
\end{equation}
\end{enumerate}
Steps 2-3 are repeated until the assignments no longer change. 
Figure~\ref{fig:LF-algo} shows an illustrative example with $n=9$ synthetic bivariate observations where the Lloyd-Forgy algorithm converges in three iterations.
The number of iterations before convergence depends on the initial mean values chosen.
In this example, the achieved local optimum is also the global optimum.

\medskip

\paragraph{\bf Hierarchical clustering}
Hierarchical methods are unmonitored clustering based on tree representations. 
The top of the tree is a single cluster with all observations while each observation is a cluster in the bottom of the tree (the leaves). 
Hierarchical methods are based on a distance metric (e.g., Euclidean, Manhattan, Mahalanobis, etc.), and a linkage between two clusters (centroid distance, min/max distance, etc.) from which we can build an objective function. 
In general, the problem of finding the optimal hierarchical clustering, i.e., the one that minimises a global objective function is NP-hard. 
This means that as the size of the data grows, finding the exact optimal hierarchical clustering for large datasets becomes computationally infeasible because the number of possible hierarchical clustering grows exponentially with the observation number.

\begin{figure}[!ht]
    \centering
    \includegraphics[width=.8\textwidth]{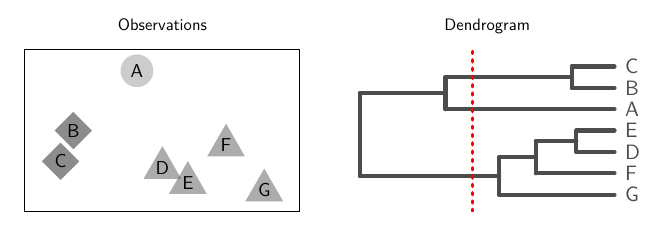}
    \caption{Seven bivariate observations (left panel) and their dendogram representation from an agglomerative clustering method with Euclidean distance and centroid linkage, with a section for three clusters: $\{A\}$, $\{B,C\}$ and $\{D,E,F,G\}$.}
    \label{fig:Dendogram}
\end{figure}

Agglomerative and divisive methods are efficient heuristic algorithms that produce good hierarchical clustering in practice, but they do not guarantee the globally optimal solution due to the NP-hardness of the problem.
Agglomerative clustering methods (ACM) are bottom-up approaches, in contrast to top-down divisive methods. 
In agglomerative methods, each observation is initially a cluster.
ACM consists of repeatedly merging the two closest clusters until the number of clusters is greater than one.
A dendrogram is a common graphical representation of hierarchical clustering. 
Dendograms are trees with the linkage distances in the $x$ coordinate (branches) and observations in the $y$ coordinate (leaves). 
The section of the dendogram allows the number of clusters and the members of each cluster  to be determined. 
A basic ACM with Euclidean distance and centroid linkage for seven synthetic bivariate observations and its dendogram representation is given in Figure~\ref{fig:Dendogram}. 
Here the shape of the dendogram suggests a configuration with two clusters: $\{A,B,C\}$ and $\{D,E,F,G\}$, or with three: $\{A\}$, $\{B,C\}$ and $\{D,E,F,G\}$, as shown in the figure.

\medskip

\paragraph{\bf Density-based clustering}
Density-based clustering techniques are non-parametric methods based on empirical data probability density functions, typically using kernel density estimates.
These algorithms are designed to identify regions where data points are densely packed and separate them from regions of lower density including noise and outliers. 
Density-based clustering problems are considered NP-hard in the general case, especially when the goal is to find the globally optimal clustering. 
However, heuristic methods offer practical solutions without necessarily finding the globally optimal clustering. 
Popular density-based non-parametric clustering heuristics were developed in the late 1990s, including mean shift \cite{cheng1995mean}, DBSCAN (Density-Based Spatial Clustering of Applications with Noise) \cite{ester1996density} or OPTICS (Ordering Points to Identify the Clustering Structure) \cite{ankerst1999optics} algorithms. 

The mean shift algorithm iteratively shifts points towards the mode (peak) of a density function until convergence \cite{cheng1995mean}. 
Clusters form around these modes. 
A bandwidth parameter controls the size of the window used to shift points.
Similarly, data points are clustered based on the density of points in a neighbourhood with the DBSCAN algorithm \cite{ester1996density}. 
This technique is controlled by two parameters: the neighbourhood search radius and the minimum number of points required to form a dense region (i.e., a cluster). 
The OPTICS algorithm is an extension of DBSCAN where the radius for neighbourhood radius is adaptive \cite{ankerst1999optics}. 
Oppositely to DBSCAN, OPTICS allows detecting meaningful clusters in data of varying density.

Unlike $K$-means and, to a lesser extent depending on the metric and linkage used, hierarchical algorithms which mainly identify circular clusters, density-based clustering methods are designed to find clusters of arbitrary shape, including non-convex clusters\footnote{A comparison of $K$-means, AHM and mean-shift clustering algorithms on synthetic bivariate datasets including circular and non-convex clusters can be computed online at\\ \href{https://colab.research.google.com/drive/1vVzTL\_Cf4kUUnbSpAZ4vVmjPKlgAljsC?usp=sharing}{\texttt{colab.research.google.com/drive/1vVzTL\_Cf4kUUnbSpAZ4vVmjPKlgAljsC}}}.
In addition, density-based clustering algorithms do not require the number of clusters to be defined. 
However, density-based models are generally sensitive to parameter values, especially the bandwidth and clustering threshold.
\medskip

\paragraph{\bf Parametric clustering}
Parametric clustering refers to a type of clustering where the data is modelled using a fixed number of parameters, such as the number of clusters or parameters related to the shape of the clusters.
Often the data are assumed to follow a particular mixture distribution model, typically multivariate Gaussian distributions. 
The clusters may have a particular shape, like spherical or ellipsoidal, depending on the parametric model.
The main objective is to fit the parameters of the model to the data in a way that best describes the underlying clusters.

For a Gaussian mixture model with multivariate normal probability density function $f$, the parameters $\theta=(\mu_k,\sigma_k,\pi_k,k=1,\dots,K)$ are, besides the number of clusters $K$, the expected values $\mu_k$ and standard deviations $\sigma_k$ as well as the proportions of observations per cluster $\pi_k$. 
The likelihood of the model is
\begin{equation}
    \mathcal L_\theta(X)=\prod_{i=1}^n \sum_{k=1}^K \pi_kf(x_i,\mu_k,\sigma_k).
\end{equation}
Local maxima can be obtained using the expectation-maximisation heuristic \cite{meng1997algorithm}, variational Bayesian methods, or randomised algorithms such as Gibbs sampling.
In addition, the number of clusters can be selected using information criteria such as AIC, BIC or likelihood ratio.

\subsection{Supervised learning methods}
\label{subsec:Supervised}

In the context of supervised learning methods the multivariate dataset reads
\begin{equation}
X=\begin{bmatrix} x_1^1&x_1^2&\ldots&x_1^p\\[1mm]
x_2^1&x_2^2&\ldots&x_2^p\\
\vdots&\vdots&&\vdots\\
x_n^1&x_n^2&\ldots&x_n^p
\end{bmatrix}\in\mathbb R^{n\times p},\qquad 
Y=\begin{bmatrix} y_1\\[1mm]
y_2\\
\vdots\\
y_n
\end{bmatrix}\in\mathbb R^{n}.
\end{equation}
The observations $X$ are the set of explanatory variables (sometimes called independent variables, predictor variables, regressors, covariates, features, or input variables), while $Y$ is the variable to be explained (sometimes called the dependent variable, response variable, predicted variable, or output variable), which we assume to be univariate for simplicity. Typically, the data are sensor measurements of various physical quantities such as temperature, pressure, vibration, speed (e.g. rotation), noise, etc. of a machine, and $Y$ is the operational state of the machine or its remaining useful life. 

Supervised learning methods are algorithms that aim to predict $Y$ from $X$. 
To this end, the parameters of the algorithm are inferred from the data, hence the learning process.
Classification methods refer to the case where $Y$ is discrete or categorical, e.g. a class such as the qualitative state of the machine. 
While regression methods refer to the prediction of continuous numeric quantities, typically remaining useful life.
Machine learning methods refer to (nonlinear) algorithms that are flexible enough to be used for both classification and continuous prediction.

\subsubsection{Multivariate regression techniques}
\label{subsubsec:Reg}
\begin{figure}[!ht]
\vspace{-5mm}
    \centering
    \includegraphics[width=.52\textwidth]{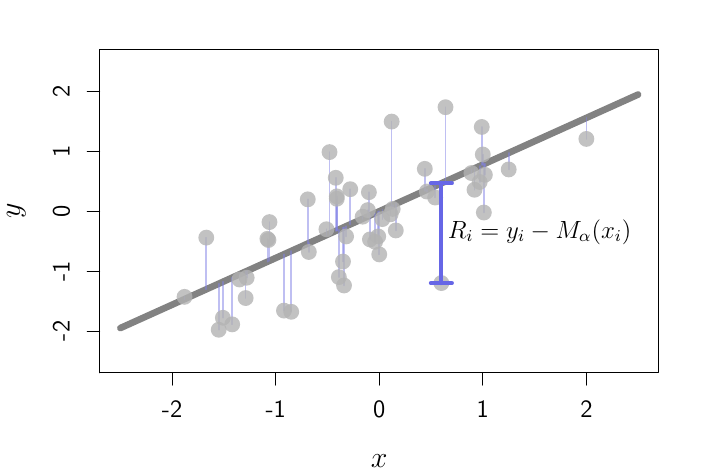}$\quad$
    \caption{Illustrative scheme of a univariate linear regression model. The parameters $\alpha$ are calibrated by minimising the squared difference between the model $M_\alpha(x_i)$ and the observation $y_i$. This corresponds to minimising the variability of the residuals $R_i=y_i-M_\alpha(x_i)$.\vspace{.5cm} }
    \label{fig:LinReg}
\end{figure}

We focus first on regression methods where the variable to be predicted is numerical and continuous.
The algorithms are models given by
\begin{equation}
    y_i=M_\alpha(x_i)+\sigma\varepsilon_i,
\end{equation}
where $\alpha$ is a set of parameters (coefficients) to be calibrated and $\varepsilon_i$ is a centred random variable with amplitude $\sigma$ corresponding to the unexplained part of the models. 
For instance, a linear model reads
\begin{equation}
    M_\alpha(x_i)=\alpha_0+\alpha_1x^1_i+\dots+\alpha_nx^n_i.
\end{equation}
The objective is to calibrate the parameters in order to fit the data as well as possible, within the limits of the model's shape.
In regressive models, the coefficients are estimated by minimising the variability of the residuals $R_i(\alpha)=y_i-M_\alpha(x_i)$, i.e. the difference between the model and the observation (see Figure~\ref{fig:LinReg}).
This is
\begin{equation}
    \tilde\alpha=\text{arg}\max_\alpha \sum_{i=1}^n \big(y_i-M_\alpha(x_i)\big)^2.
\end{equation}

For linear models, it is easy to check that 
\begin{equation}
    \tilde\alpha=\big(X^\top X\big)^{-1}X^\top Y,
\end{equation}
while the unbiased estimate of the noise amplitude is given by
\begin{equation}
    \tilde\sigma^2=\frac1{n-p-1}\sum_{i=1}^nR^2_i(\tilde\alpha).
\end{equation}
For example, for a univariate linear model $M(x)=ax+b$, $a,b\in\mathbb R$, the least squares estimate is given by $\tilde a=\text{covar}(x,y)/\text{var}(x)$ and $\tilde b=\bar y-\tilde a\bar x$. 
This least squares estimate maximises the likelihood if the residuals are independent and identically normally distributed. 
It is in general a good estimate for compact distributions, but is sensitive to extreme values and outliers. 
Fisher tests can be used to assess the influence of a single feature, a group of features, or the linear model as a whole.
Nonlinear models can be handled by using invertible nonlinear transformations. 
For instance, linear regression of a log/log transformation $\hat x=\log(x)$ and $\hat y=\log(y)$ leads to the exponential model 
\begin{equation}
    M_\alpha(x)=e^{\alpha_0}\cdot(x^1)^{\alpha_1}\ldots(x^p)^{\alpha_p}.
\end{equation}
The logarithmic model 
\begin{equation}
    M_\alpha(x)=\log(\alpha_0+\alpha_1x^1+\ldots+\alpha_px^p)
\end{equation}
can be treated with the exponential transformation $\hat y=e^y$.
More generally, regressions for differentiable nonlinear models are based on linearisation and iterative heuristics, such as the Gauß-Newton algorithm, which converge to local optima.

The quality of the fit can be assessed by partitioning the total variability $SST$ into the variability explained by the model $SSM$ and the variability of the residuals $SSR$, where
\begin{equation*}
SST=\sum_{i=1}^n\big(y_i-\bar y\big)^2,\quad SSM=\sum_{i=1}^n\big(\bar M-M_{\tilde\alpha}(x_i)\big)^2,\quad SSR=\sum_{i=1}^n\big(y_i-M_{\tilde\alpha}(x_i)\big)^2.
\end{equation*}
Indeed, we have $SST=SSM+SSR$ when the residuals are centred.
Therefore, minimising the variability of the residuals leads back to maximising the variability of the model.
In addition, the proportion of variability explained 
\begin{equation}
R^2=\frac{SSM}{SST}=1-\frac{SSR}{SST},
\end{equation}
known as the coefficient of determination, can be used to quantify the quality of the fit. 
We have $R^2\in[0,1]$ for centred residuals. 
The coefficient of determination is equal to one if the total variance is recovered by the model, while it is equal to zero if the variability of the model is equal to zero.
Note that for univariate data, $R^2=\text{cor}^2(x,y)$.

\subsubsection{Classification techniques}
\label{subsubsec:Classif}
In this section, we focus on statistical classification methods where the variable to be predicted is discrete.
\medskip

\paragraph{\bf Logistic regression}
The logistic regression is a binary classification algorithm for which $Y\in\{0,1\}^n$. 
The first mention of the logistical model can be traced back to the work of Berkson in 1944 \cite{berkson1944application}.
In the logistic model, the membership of an observation to a class is quantified with the help of the sigmoid (logistic) function
\begin{equation}
    \tilde y_i=\pi_i(\alpha)=\frac{1}{1+\exp\big(-(\alpha_0+\alpha_1x_i^1+\ldots+\alpha_px_i^p)\big)}.
\end{equation}
This quantity lies between 0 and 1 with a relatively high transition speed: 
the logistic function tends quickly to zero if $\alpha_0+\alpha_1x_i^1+\ldots+\alpha_px_i^p<\!\!<0$ while it fast tends to one if $\alpha_0+\alpha_1x_i^1+\ldots+\alpha_px_i^p>\!\!>0$.
The scalars $(\alpha_0, \alpha_1, \dots , \alpha_p)$ are the $p + 1$ parameters (coefficients) of the logistic model to be estimated and interpreted. 
Note that binary variables can be generalised to categorical variables when there are more than two possible values, and the binary logistic regression can be generalised to multinomial logistic regression.

The parameters of the binary logistic regression can be estimated by maximising the likelihood 
\begin{equation}
    \mathcal L(\alpha)=\prod_{i=1}^n \pi_i(\alpha)^{y_i}(1-\pi_i(\alpha))^{1-y_i}
\end{equation}
using heuristics, typically gradient methods.
In practice, regularisation is often added to the optimisation problem to prevent overfitting, especially when dealing with high dimensional data.
The parameters (coefficients) in logistic regression reflect the log-odds of the outcome as a function of the input features. The intercept $\alpha_0$ represents the log-odds of the outcome when all the input features are zero.
If $\alpha_0$ is large and positive, it indicates that the predicted probability of the outcome is high when all feature values are zero, and vice versa. 
For a positive coefficient $\alpha_j$, the feature $x^j$ increases the likelihood of the outcome $y=1$. The higher the value of the feature, the higher the probability of the positive outcome. The opposite happens for negative coefficients. 
To make the coefficients easier to interpret, they are often exponentiated to convert them from log-odds to odds ratios. 
\medskip

\paragraph{\bf Linear discriminant analysis}
Linear discriminant analysis consists of a linear transformation of the explanatory variables to maximise the inter-class variability or, equivalently, to maximise the intra-class variability. 
The principle is similar to that of the principal component analysis. 
The optimal linear transformation can be obtained explicitly by matrix calculus, using the variance/covariance matrix. 
Linear discriminant analysis deals with binary output data with two classes and can be extended to multi-class analysis \cite{rao1948utilization}.
In fine, the binary classification resumes to a linear discriminant
\begin{equation}
    D(\alpha)=\alpha_0+\alpha_1x^1+\dots+\alpha_px^p. 
\end{equation}
In the standard score transformation of the data, the belonging of an observation to one of the two classes is given by the sign of the discriminant. 
As in  logistic regression, the coefficient estimates allow the influence of a given feature on the class to be interpreted. 
Note that the linear discriminant analysis is much more sensitive to extreme values and outliers than the logistic regression.
\medskip 

\paragraph{\bf Naive Bayes algorithm}
Unlike previous logistic and discriminant classifiers, the naive Bayes algorithm is directly designed for multi-class problems.
The classifier is based on the Bayes theorem
\begin{equation}\label{eq:Bayes}
    \mathbb P(Y|X)=\frac{\mathbb P(Y)\mathbb P(X|Y)}{\mathbb P(X)},\qquad \mathbb P(X)>0,
\end{equation}
where $Y$ is the (latent) variable to be predicted and $X$ the (observable) explanatory variables. 
A natural estimation for $Y$ is the value that maximise its probability $\mathbb P(Y|X)$ conditionally to the observations. 
In \eqref{eq:Bayes}, $\mathbb P(X)$ is a constant that does not depend on $Y$ while $\mathbb P(Y)$ can be estimated using empirical means $\tilde p_k$.
The objective is therefore to estimate the conditional probability $\mathbb P(X|Y)$. 
In the naive Bayes algorithm, the features are assumed to be independent and normally distribution so that the prediction is given by 
\begin{equation}
    \tilde y_i=\text{arg}\max_k ~\tilde p_k\prod_{j=1}^p f (x_i^j,\mu_j^k,\sigma_j^k)
\end{equation}
with $f$ the normal probability density function. The $2Kp$ parameters $(\mu_j^k,\sigma_j^k)$ can be estimated by maximum likelihood, i.e., by least square estimates over the classes as the distributions are assumed normal and independent.
Similarly to the linear regression, the Bayes classifier is sensitive to extreme values.
\medskip

\paragraph{\bf Decision tree}
Logistic regression, linear discriminant analysis and naive Bayes classifier are linear algorithms providing limited classification for groups whose shape does not allow differentiation by hyperplanes. 
In contrast, decision trees are based on recursive binary partition rules that can recover nonlinear features. 
The goal is to split the data in a way that results in the most homogeneous subgroups (in terms of the target variable).
As Bayes classifier, they are natural designed for multi-class problem and categorical data.
Furthermore, they are easy to interpret and visualise using dendograms, especially for small trees, and don’t assume any specific data distribution, making them flexible for various tasks.
Decision trees also naturally rank the importance of features based on how often and effectively they are used to split the data. 
However, they are due to splitting iteration computationally expensive, can lead to overfitting (that can be mitigated by pruning the tree or setting stopping criteria) and can prove unstable (i.e. sensitive to particular observations).

\subsubsection{Machine learning techniques}
\label{subsubsec:ML}
Machine learning techniques are statistical algorithms that can learn from data and generalise to make predictions.
However, in contrast to the classical statistical models described above, machine learning algorithms are almost not constrained by the shape of the model \cite{breiman2001statistical,saporta2008models}.
In fact, they contain many more coefficients, giving them great malleability and allowing them to approximate a wide variety of relationships, including complex nonlinear and categorical  ones.
The learning mechanisms shape the algorithms through the data, empirically identifying the underlying mechanisms at work. 
Unlike traditional statistical approaches, the predictive power comes from the data rather than the shape of the model. 
In this sense, machine learning approaches are not based on a theory used to formulate a model.
They have no modelling bias and can be applied to a wide range of problems, including classification and regression. 
However, although their predictive ability can be much more accurate than classical statistical models, they do not allow the influence of the features to be understood directly by interpreting the parameter values, as can be done with models. 
For this reason, they are often referred to in the literature as \emph{black boxes}.

The learning processes that allow the data to be fitted and the coefficients to be calibrated are fundamental to machine learning algorithms.
This is the model training phase, formulated as a multidimensional optimisation problem. 
The objective is to minimise a prediction error function, which can be a squared error or an entropy function.
The size of the problem, i.e. the number of coefficients to be calibrated, is very large (hundreds, even thousands if not more), so training is performed systematically using iterative heuristics, typically backpropagation methods, which compute the gradient of a loss function with respect to the network weights. 
Tuning the complexity of the algorithms and their hyperparametrisation is, in the absence of theory, an open problem that needs to be specified according to the input data. 
Hyperparameters are, for example, the number of layers and neurons in artificial neural networks, the number of support vectors in support vector machines or the number of decision trees in random forests.
Although theoretical tools exist for defining the optimal complexity of algorithms, such as the Vapnik-Chervonenkis dimension, they are often difficult to apply.
In practice, determining the optimal complexity of algorithms is usually determined empirically by partitioning the data using cross-validation techniques, where part of the data is used to train the algorithms while the remaining part is used to test the predictions on new data.
These operations can be repeated by randomly partitioning the data in training and testing to quantify the uncertainties of the prediction in so-called bootstrapping techniques. 
If the training error function decreases systematically as the complexity of the algorithms increases, the test error decreases before reaching a minimum and then increases. 
The minimum corresponds to optimal complexity, the decreasing part to insufficient complexity, known as underfitting, and the increasing part to overparametrisation, known as overfitting, which prevents generalisation. 
\medskip

\begin{figure}[!ht]
    \centering
    \includegraphics[width=.48\textwidth]{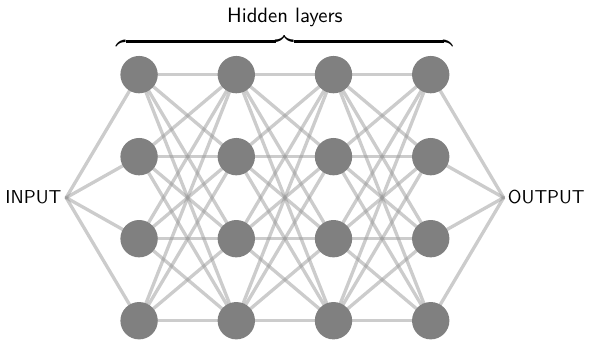}\hfill\includegraphics[width=.48\textwidth]{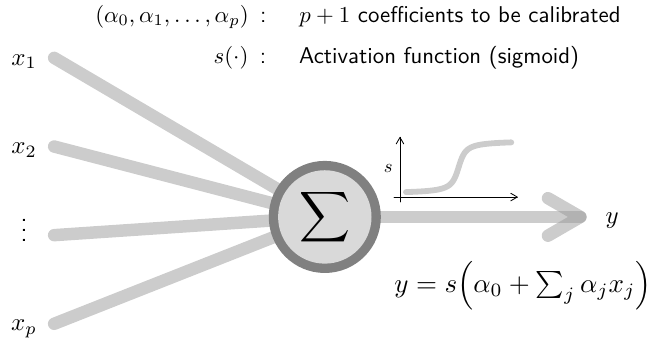}
    \caption{Scheme of an artificial feed-forward neural network. Left panel: Layer structure of fully connected neurons. Right panel: Activation operation of a neuron.}
    \label{fig:NN}
\end{figure}

\paragraph{\bf Artificial neural network}
Inspired by the way a brain works, artificial neural networks (ANN) are networks of fully connected neurons organised in layers (see Figure~\ref{fig:NN}, left panel). 
The first artificial neural network algorithms can be traced back to the pioneering work of Lettvin et al. on the eye and brain  of a frog \cite{lettvin1959frog}.
Each neuron is a activation function (e.g., a sigmoid) on a linear combination of the input, calibrated by weighting coefficients. 
Since neurons are fully connected, the coefficient number is the dimension of the input plus one times the number of neurons. 
Using a logistic activation function, a single neuron network is simply a logistic model. 

The malleability of the algorithm arises through the superposition of many layers and neurons. 
In fact, any relationship can approximated using a neural network, provided the network is sufficiently complex. 
ANNs have found applications in many disciplines, including regression analysis and classification. 
Different types of architecture are used in practice, depending on the discipline. 
The simplest networks are feed-forward networks in which information propagates in one direction only. 
Deep neural networks
Recursive approaches include loops that can operate at different scales, as in long-short term memory (LSTM) networks. 
These approaches are well adapted to the analysis of time series. 
Convolutional neural networks consist of a succession of subsampling of overlapping inputs and convolution, which is particularly well suited to image processing. 
In addition, networks are said to be deep when the number of layers (and neurons) is large.

\medskip 

\paragraph{\bf Support vector machine}
Artifical neural networks consists of a tangle of nonlinear activation functions where the phase of training requires using approximation heuristics. 
In contrast, support vector machines (SVM) rely on combinations of optimal linear hyperplane in a high dimensional space, making them analytically tractable. 
The optimisation is carried out by maximising margins between the closest data points of different classes (i.e., the hyperplane and support vectors). 
It focuses on the points that are hardest to predict.
This makes SVMs to be resilient to noisy data, extreme values and outliers unlike classical linear statistical models such as linear discriminant, naive Bayes classifier, or linear regression.

As ANNs, SVMs can be used for both classification and regression \cite{drucker1996support}, although the algorithm was initially designed for classification \cite{boser1992training}. 
Indeed, the used of multiple hyperplane make it possible the fitting of various relationships, whether nonlinear or categorical. 
In addition, similarly to nonlinear regression models, the input space can be rescaled using nonlinear transformations. 
This allows the algorithms to fit maximum-margin hyperplane in a transformed feature space that is nonlinear in the original input space and so addressing complex data structures.
Another approach to treat data that is not linearly separable, is to use a technique called the kernel trick projecting the data into higher-dimensional space, where it becomes linearly separable  \cite{boser1992training}. 
Common kernels include polynomial, Gaussian and radial basis function kernels.
SVMs are especially effective in high-dimensional spaces and when the number of dimensions exceeds the number of samples. They are also memory efficient, as it uses only the support vectors.
However, SVMs can be less effective with large datasets because training complexity can become high. 
In addition choosing the right kernel and hyperparameters (e.g., number of support vectors) can be tricky.
\medskip
    
\paragraph{\bf Ensemble learning}
Ensemble methods in machine learning are techniques that combine the predictions of multiple algorithms (often referred to as "weak learners" or "base learners") to produce a more accurate and robust prediction than any single algorithm. 
The key idea behind ensemble methods is that by leveraging the diversity of algorithms, the ensemble can correct for the weaknesses of individual models and improve overall performance.
Indeed different models may capture different patterns or aspects of the data, and combining them tends to compensate the weakness and reduce errors.
The main types of ensemble methods include bagging methods (bootstrap aggregating), stacking (or stacked generalisation) and boosting.
\begin{itemize}
    \item {\bf Bagging methods} \cite{breiman1996bagging} reduce variance by training multiple instances of the same algorithm on different random subsets (bootstrapped samples) of the training data, and then averaging their predictions.
    Random forest is a popular bagging algorithm that trains multiple decision trees and averages their results (for regression) or votes on them (for classification).
    \item {\bf Stacking methods} \cite{wolpert1992stacked,breiman1996stacked} involve training several basic algorithms and then using a \emph{meta-heuristic} to combine their predictions. The meta-heuristic learns how best to aggregate the outputs of the base models.
    A common setup is to use random forests, SVMs, and neural networks as base models, and then train a linear regression or another model to combine their outputs.
    \item {\bf Boosting methods} aim to reduce bias by training algorithms sequentially, focusing on the errors made by the previous algorithms, and progressively improving the prediction.
    AdaBoost (Adaptive Boosting) \cite{freund1995desicion} and Gradient Boosting \cite{friedman2001greedy} are common boosting algorithms. In AdaBoost, misclassified data points are given higher weights in each iteration, while Gradient Boosting iteratively optimises the loss function.
\end{itemize}

Ensemble learning algorithms are particularly suited to the analysis of complex multivariate datasets. 
They are widely used in both academic research and industrial applications.
Ensemble methods, such as bagging, can help reduce overfitting by averaging out the biases of individual models, leading to a more general and robust prediction algorithm.
Boosting techniques can reduce bias by focusing on correcting the errors made by previous algorithms. 
These two strategies help in balancing the trade-off between bias and variance in machine learning models.
However, training multiple algorithms requires more computational resources and time than training a single algorithm.
In addition, ensemble methods, combining many algorithms, are difficult to interpret. 

\medskip
Applications of different algorithms for the prediction of pedestrian speeds in single-file movement is given for illustration in Appendix~4\footnote{Multi-algorithm predictions of pedestrian speeds can be computed online at\\ \href{https://colab.research.google.com/drive/1zGHJQbosu-gZPB8HUaf0GVHSJb_hdcbG?usp=sharing}{\texttt{colab.research.google.com/drive/1zGHJQbosu-gZPB8HUaf0GVHSJb\_hdcbG}}\\
Additionally, multi-algorithm predictions for other datasets including discrete lane-changing maneuvers on highways and RUL aircraft engine forecasting can be computed online at\\
\href{https://colab.research.google.com/drive/1R_p1oJfP_PRke9eYXxRCpknQXTiUuh8W?usp=sharing}{\texttt{colab.research.google.com/drive/1R\_p1oJfP\_PRke9eYXxRCpknQXTiUuh8W}}\\
\href{https://colab.research.google.com/drive/1KSbi3Lms9hSTx-pnceS0SjltV1XTm7ZU?usp=sharing}{\texttt{colab.research.google.com/drive/1KSbi3Lms9hSTx-pnceS0SjltV1XTm7ZU}}}.

\bibliographystyle{abbrv}
\bibliography{refs}

\bigskip\bigskip

\section*{Appendix}
\addcontentsline{toc}{section}{Appendix}

\subsection*{Appendix 1: Examples of fail-operational M-out-of-N architectures}\label{AppA}
\addcontentsline{toc}{subsection}{Examples of fail-operational M-out-of-N architectures}

We detail two examples of fail-operational M-out-of-N electronic control unit architectures in this appendix: a \emph{2-out-of-2 Dual Fail Safe} (DFS) architecture and a \emph{Trimodular 2-ouf-of-3} (TMR) architecture.

\begin{figure}[!ht]
    \centering
    \begin{subfigure}[b]{0.48\textwidth}
        \centering
        \def\svgwidth{\textwidth}    
        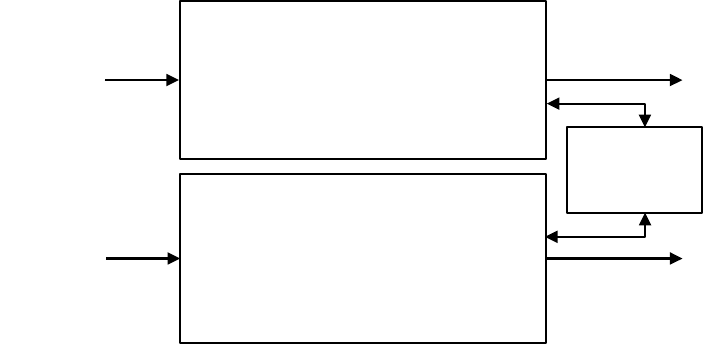 
        \caption{2-out-of-2 Dual Fail Safe (DFS)}
        \label{fig:image1}
    \end{subfigure}
    \hfill
    \begin{subfigure}[b]{0.48\textwidth}
        \centering
        \def\svgwidth{\textwidth}    
        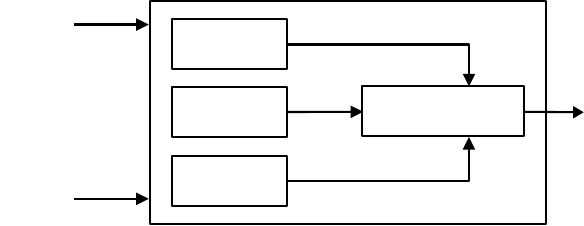 
        \caption{Trimodular 2-ouf-of-3 (TMR)}
        \label{fig:image2}
    \end{subfigure}
    \caption{Fail-operational M-out-of-N electronic control unit (ECU) architectures~\cite{Julitz.2022, Julitz.2023}}
    \label{fig:ECU}
\end{figure}

The \emph{2-out-of-3 (TMR)} architecture involves a redundancy scheme where three independent processing paths receive sensor information from the same sources (e.g., camera, radar, lidar), see Fig.~\ref{fig:image2}. A majority decision is made between the three independent processing paths, and the system continues to operate as long as at least two of the three paths remain functional, ensuring fail-operational behavior. However, a drawback of the 2oo3 architecture is that all three processing paths typically reside within the same ECU. This makes the system more vulnerable to common-cause failures, such as hardware defects or environmental impacts affecting the entire ECU.

In contrast, the \emph{2-out-of-2 Dual Fail Safe (DFS)} architecture consists of two independent fail-safe subsystems distributed across two separate ECUs, see Fig.~\ref{fig:image1}. These subsystems communicate with each other in isolation, and in the event of a fault, one ECU can take over the functionality of the other. A key advantage of the 2oo2 architecture is the greater ability to implement separation and diversification, as the two ECUs can be physically isolated and use different technologies, which significantly reduces the risk of common-cause failures. However, a drawback is that once one ECU fails, the system is no longer fail-operational and can only maintain fail-safe behavior, requiring restoration of the faulty ECU for full functionality.

\newpage
\subsection*{Appendix 2: Markov transition diagram of a 2-out-of-3 system}\label{AppB}
\addcontentsline{toc}{subsection}{Markov transition diagram of a 2-out-of-3 system} 

\begin{figure}[!ht] 
    \medskip\centering                  
    \def\svgwidth{.8\textwidth}    
    \input{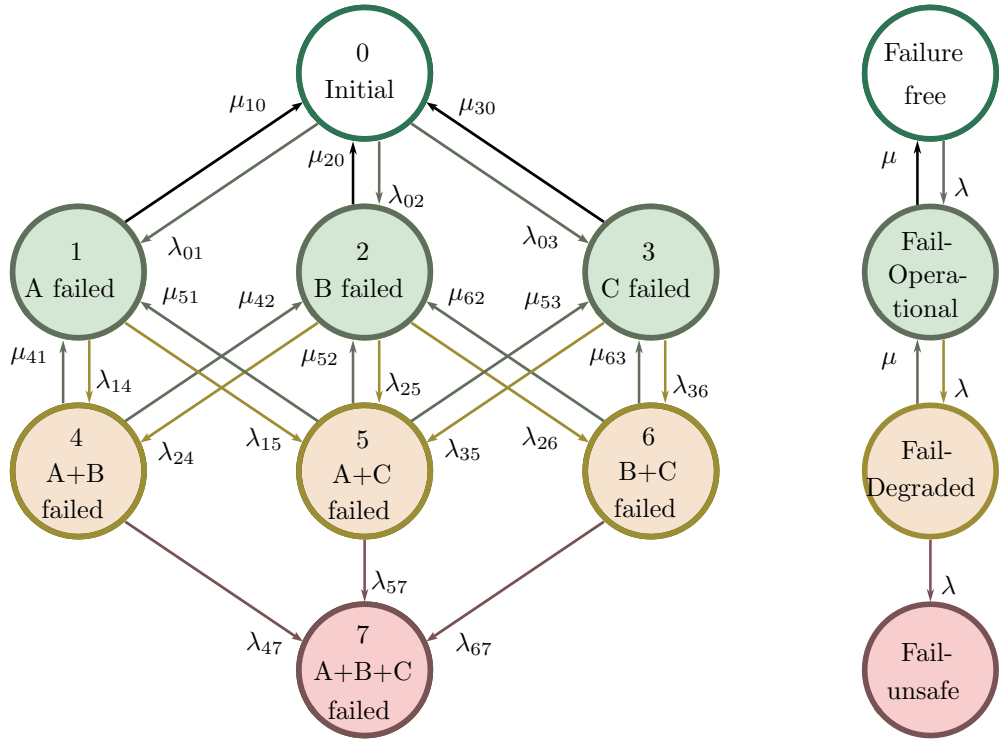}
    \caption{Markov transition diagram of a 2-out-of-3 System with 4 operational modes~\cite{Haring.2022}} 
    \label{fig:markov}          
\end{figure}

 We present in this appendix the Markov transition diagram of a 2-out-of-3 System with 4 operational modes, see Fig.~\ref{fig:markov}.
 
\subsection*{Appendix 3: Principal component analysis of banknote measurements}\label{AppC}
\addcontentsline{toc}{subsection}{Principal component analysis of banknote measurements}

\begin{figure}[!ht]
    \centering\vspace{-2mm}
    \begin{minipage}[c]{.65\textwidth}
    \centering
    $x^1$\\[-1mm]
    $\xleftrightarrow{.8\textwidth}$\\[1mm]
    \includegraphics[width=.8\textwidth]{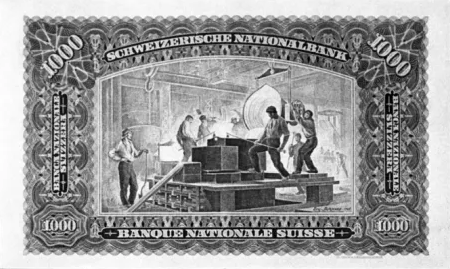}\\[-.46\textwidth]
    \hspace{-1.04\textwidth}$x^4$\\[-.048\textwidth]
    \hspace{-0.96\textwidth}\rotatebox{90}{$\xleftrightarrow{.024\textwidth}$}\\[.38\textwidth]
    \hspace{-1.04\textwidth}$x^5$\\[-.048\textwidth]
    \hspace{-.96\textwidth}\rotatebox{90}{$\xleftrightarrow{.024\textwidth}$}\\[-.275\textwidth]
    $x^2$\hspace{.93\textwidth}$x^3$\\[-.256\textwidth]
    \rotatebox{90}{$\xleftrightarrow{.46\textwidth}$}\hspace{.85\textwidth}
    \rotatebox{90}{$\xleftrightarrow{.46\textwidth}$}~\\[-.37\textwidth]
    \rotatebox{28}{$\xleftrightarrow{.5\textwidth}$}~\\[-.24\textwidth]
    \hspace{.4\textwidth}$x^6$\vspace{.3\textwidth}
    \end{minipage}\qquad\begin{minipage}[c]{.25\textwidth}
    \includegraphics[width=\textwidth]{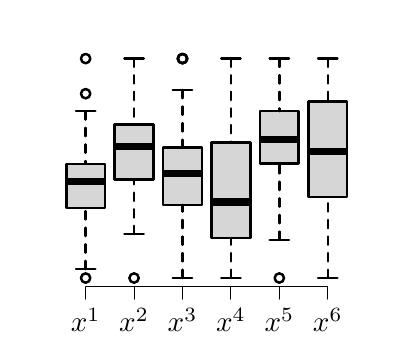}\hspace{-5mm}\\[1mm]
    \includegraphics[width=\textwidth]{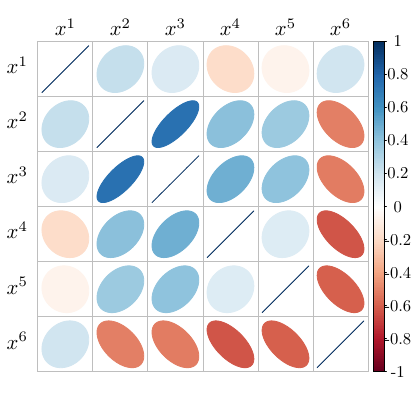}
    \end{minipage}\vspace{-1mm}
    \caption{Multivariate dataset of six measurements made on 100 genuine and 100 counterfeit old Swiss 1000 franc banknotes \cite{FluryBernhard1988Msap}. Left panel: six measurements made. Top right panel: Boxplots of the normalised measurements. Bottom right panel: Correlation table of the measurements}
    \label{fig:Banknote}
\end{figure}

As an illustrative example, we analyse the principal components of a dataset of six measurements made on 100 genuine and 100 counterfeit old Swiss 1000 franc banknotes \cite{FluryBernhard1988Msap} (see Figure~\ref{fig:Banknote})\footnote{The principal component analysis of the banknote measurements can be computed online at\\ \href{https://colab.research.google.com/drive/1D-Z-lWD_Z74Ts88EQfEjc2W1BLZBdqNw?usp=sharing}{\texttt{colab.research.google.com/drive/1D-Z-lWD\_Z74Ts88EQfEjc2W1BLZBdqNw}}}. 
The height measurements $x^2,\ldots,x^5$ are all positively correlated while the banknote width $x^1$ is more independent.
In addition, the diagonal measurement $x^6$ is negatively correlated with all the height measurements $x^2,\ldots,x^5$ and slightly positively correlated with the banknote width $x^1$.

\begin{figure}
    \centering
    \includegraphics[width=.70\textwidth]{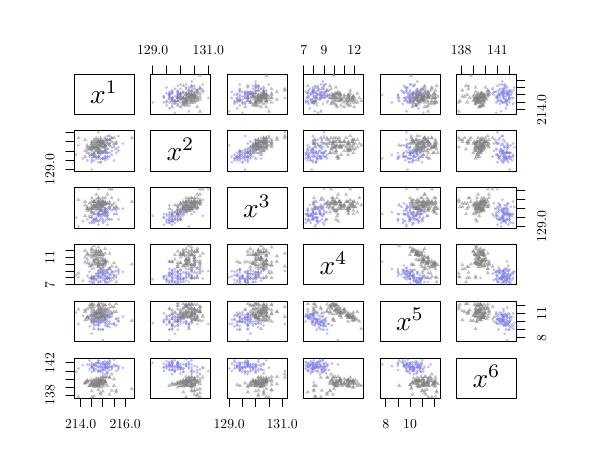}\vspace{-5mm}
    \caption{Pairwise scatterplots of the six measurements on the notes. The blue circles are the genuine banknotes while the grey triangles are the counterfeits.}
    \label{fig:Scatter-plot}
\end{figure}

\begin{figure}
    \centering
    \includegraphics[width=.37\textwidth]{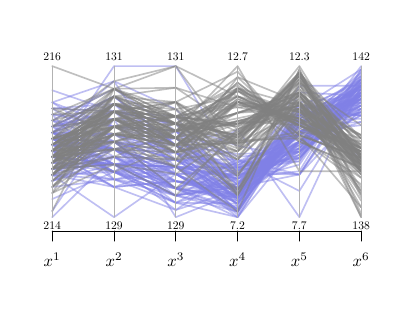}\hspace{-10mm}
    \includegraphics[width=.37\textwidth]{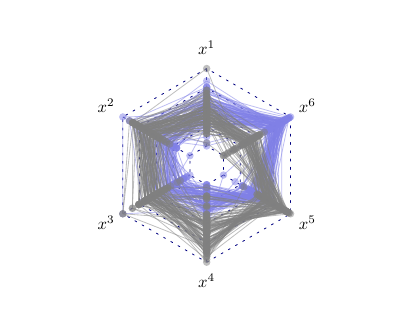}\hspace{-12mm}
    \includegraphics[width=.37\textwidth]{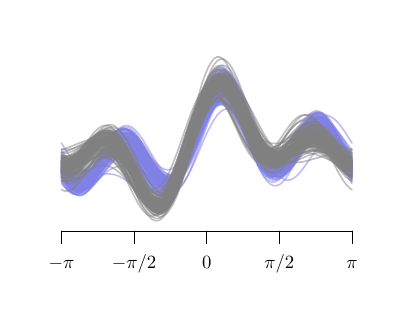}\vspace{-4mm}
    \caption{Parallel-plots (left panel), radar charts (middle panel) and Andrews' curves (right panel) for the six measurements on the notes. The blue curves are the genuine banknotes while the grey curves are the counterfeits. }
    \label{fig:PP-RC-Andrews}
\end{figure}

\begin{figure}
    \centering
    \includegraphics[width=.4\textwidth]{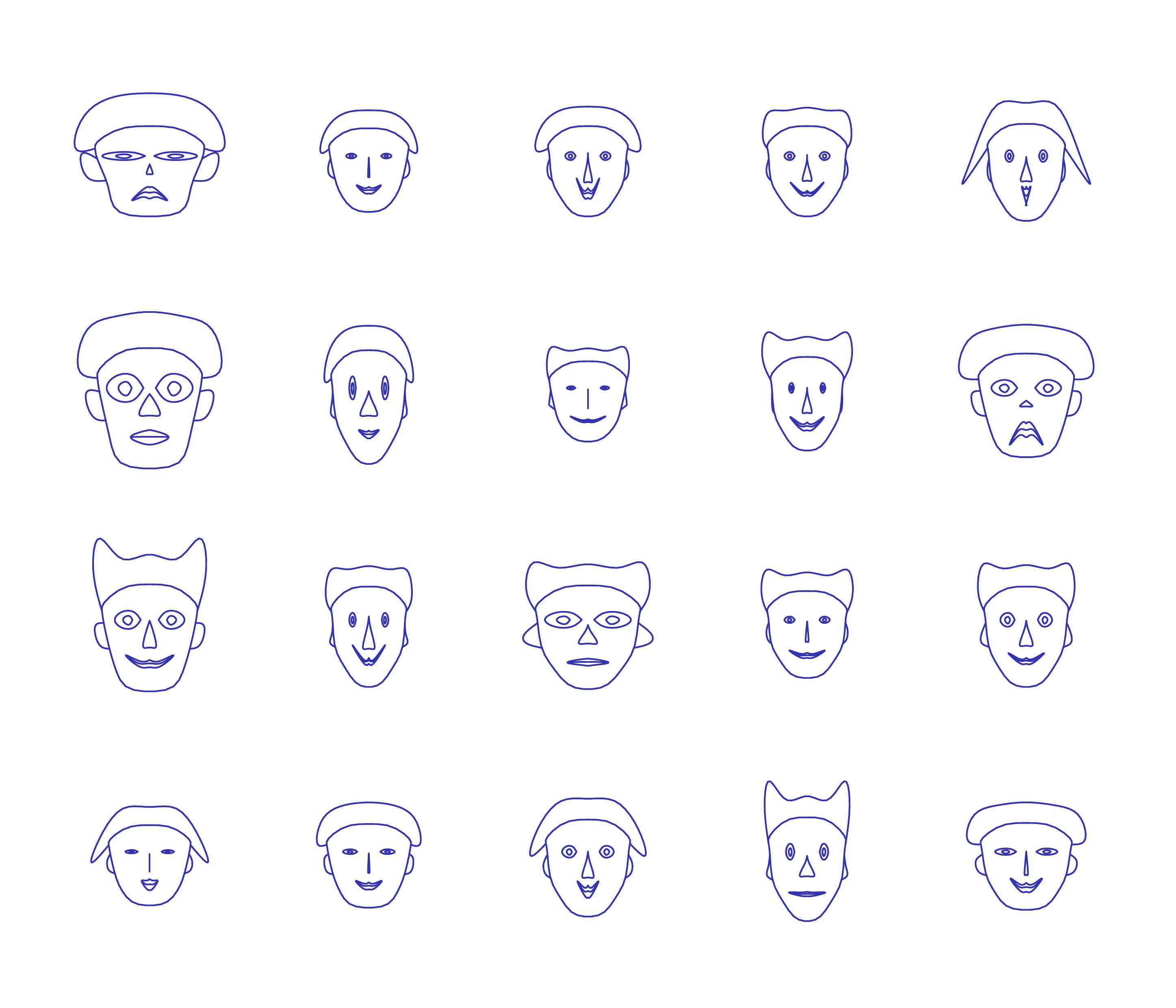}~~~~
    \includegraphics[width=.4\textwidth]{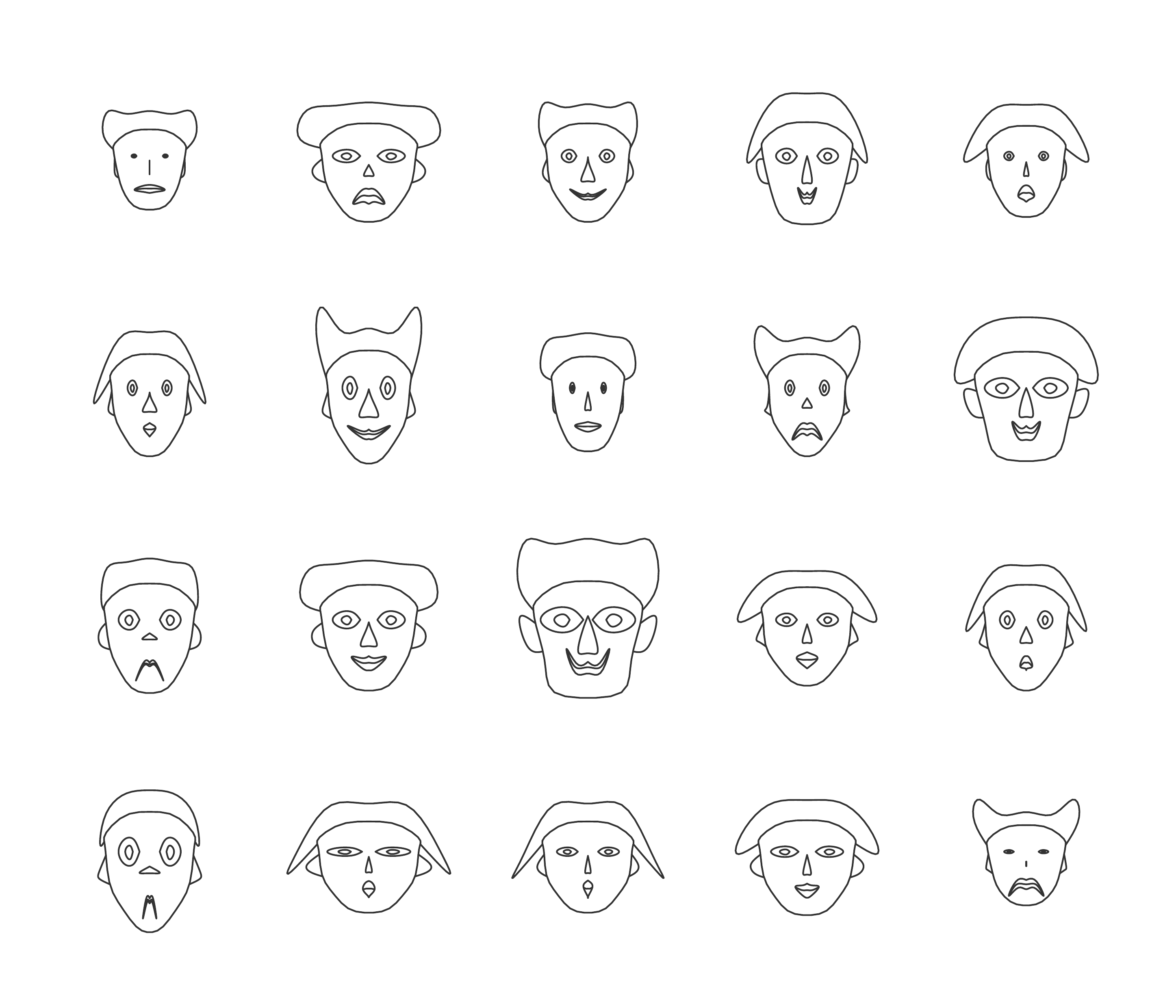}\vspace{-2mm}
    \caption{Examples of Chernoff's faces for the six measurements on the notes. The twenty faces on the left are genuine banknotes, while the twenty grey faces on the right are counterfeits.}
    \label{fig:ChernoffFace}
\end{figure}

The pairwise scatter plots of the six measurements are shown in Figure~\ref{fig:Scatter-plot}. 
The blue circles are the genuine banknotes, while the grey triangles are the counterfeits.
Two clusters of observations corresponding to the genuine and counterfeit notes can be identified, particularly using the diagonal measurement $x^6$. 
However, it is not straightforward to determine whether a given banknote is genuine or counterfeit from the scatter plots. 
Figure~\ref{fig:PP-RC-Andrews} presents some synthetic representations of the datasets using a parallel plot, a radar chart and Andrews' curves, while Figure~\ref{fig:ChernoffFace} shows Chernoff's faces of twenty genuine and twenty counterfeit banknotes, specifically in the parallel plot and the radar chart. 
Different features can be observed for genuine and counterfeit banknotes. 
However, again, it is not easy to determine from these plots whether a particular banknote is genuine or counterfeit. 

\begin{figure}[!ht]
    \centering
    \begin{minipage}[c]{.25\textwidth}
    \includegraphics[width=\textwidth]{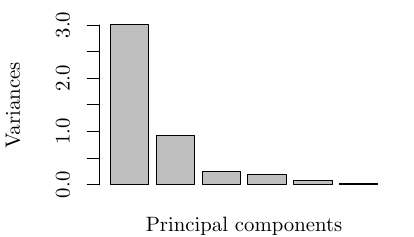}\hspace{-5mm}\\[2mm]
    \includegraphics[width=\textwidth]{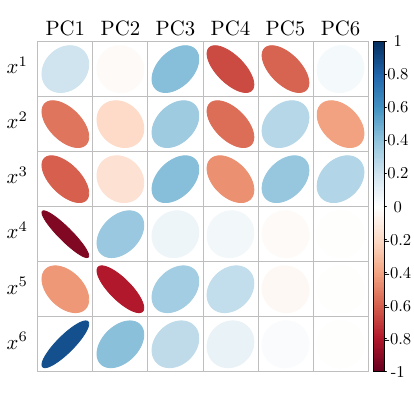}
    \end{minipage}$\qquad$\begin{minipage}[c]{.52\textwidth}
        \includegraphics[width=\textwidth]{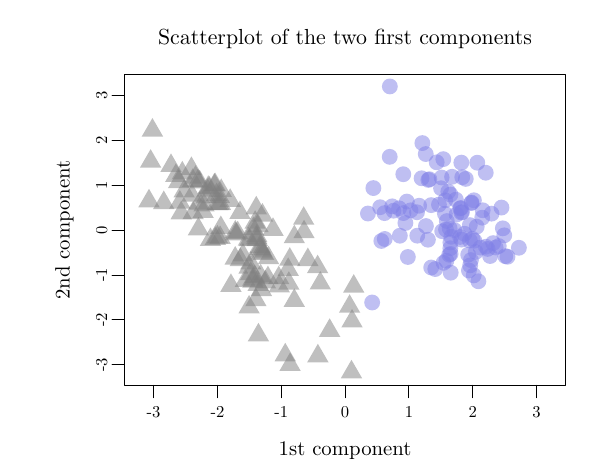}
    \end{minipage}
    \caption{Principal component analysis of the banknote dataset. Top left panels: variance per component -- the first two components contain most of the variability. Bottom left panel: Correlation table with the original data - the first two components are correlated with all the six measurements. Right panel: scatter plot of the first two components, clearly showing two groups consisting of the genuine banknotes (blue circles) and the counterfeits (grey triangles).}
    \label{fig:BanknotePCA}
\end{figure}

Figure~\ref{fig:BanknotePCA} shows the principal component analysis of the banknote dataset. 
The first two principal components contain about 90\% of the data variability (top left panel). 
Therefore, we can reasonably resume the data in two dimensions using the first two components. 
The correlation table with the original data shows that both first components are correlated with all measurements (bottom left panel). 
The first component is negatively correlated with all height measures $x^2,\ldots,x^5$ and positively correlated with the banknote width $x^1$ and the diagonal $x^6$. 
This component characterises the fact that a banknote is wide rather than long.
The second component is mainly negatively correlated with the height of the lower edge $x^5$ and positively correlated with the height of the upper edge $x^4$ and the diagonal $x^6$. 
This component rather focuses on the position of the central drawing.
Notwithstanding these interpretations, the scatter plot of the first two components (right panel) clearly shows two groups consisting of the genuine banknotes (blue circles) and the counterfeits (grey triangles). 
Most of the variability in the data is specific to the genuine and counterfeit banknotes, and the representation over the first two components captures this directly.

\subsection*{Appendix 4: Benchmarking pedestrian speed prediction algorithms}\label{AppD}
\addcontentsline{toc}{subsection}{Benchmarking pedestrian speed prediction algorithms}

\begin{figure}[!ht]
    \centering\vspace{-0mm}
    \begin{minipage}[c]{.72\textwidth}
    \includegraphics[width=\textwidth]{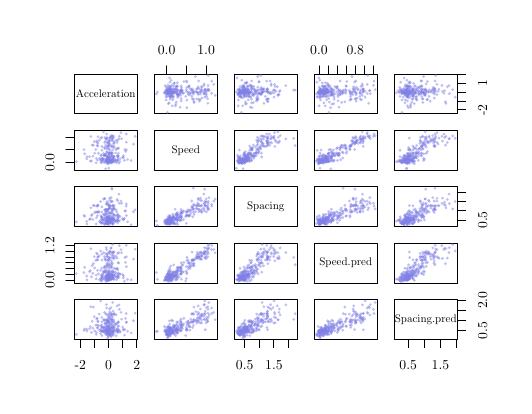}
    \end{minipage}\hfill\begin{minipage}[c]{.28\textwidth}
    \includegraphics[width=\textwidth]{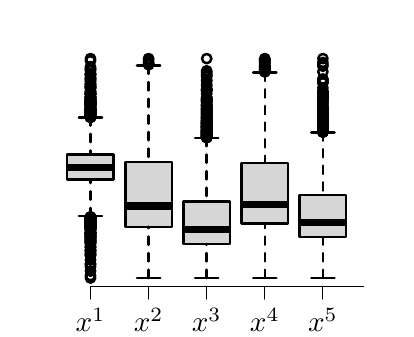}\hspace{-5mm}\\[-1mm]
    \includegraphics[width=\textwidth]{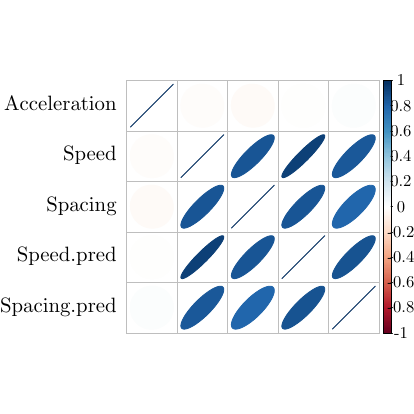}
    \end{minipage}\vspace{-7mm}
    \caption{Acceleration (in m/s$^2$), speed (in m/s), spacing (m) and speed and spacing of the neighbour in front of pedestrian in a single-file experiment. Left panel: scatterplots of the features. Top right panel: Boxplots of the normalised measurements. Bottom right panel: Correlation table.}
    \label{fig:PedData1}
\end{figure}

Pedestrian trajectory prediction is of great interest, especially for autonomous vehicles. 
In this appendix we compare the performance of different pedestrian speed prediction algorithms. 
The data come from a single file experiment where pedestrians walk in a raw on a periodic geometry without overtaking possibility\footnote{Some details on the data are available at  \href{https://ped.fz-juelich.de/da/doku.php?id=corridor2}{\texttt{https://ped.fz-juelich.de/da/doku.php?id=corridor2}}}. 
The explanatory variables are the acceleration, the spacing to the predecessor and its speed, and the spacing of the predecessor with a total of six thousand observations.
The predicted variable is the speed of the pedestrian.
A summary of the data including scatterplot, boxplot and correlation table is given in Figure~\ref{fig:PedData1}. 
Speed is correlated with all explanatory variables except acceleration. 
In addition, we can identify a nonlinear relationship between speed and distance.

\begin{figure}[!ht]
    \centering
    \includegraphics[width=.85\textwidth]{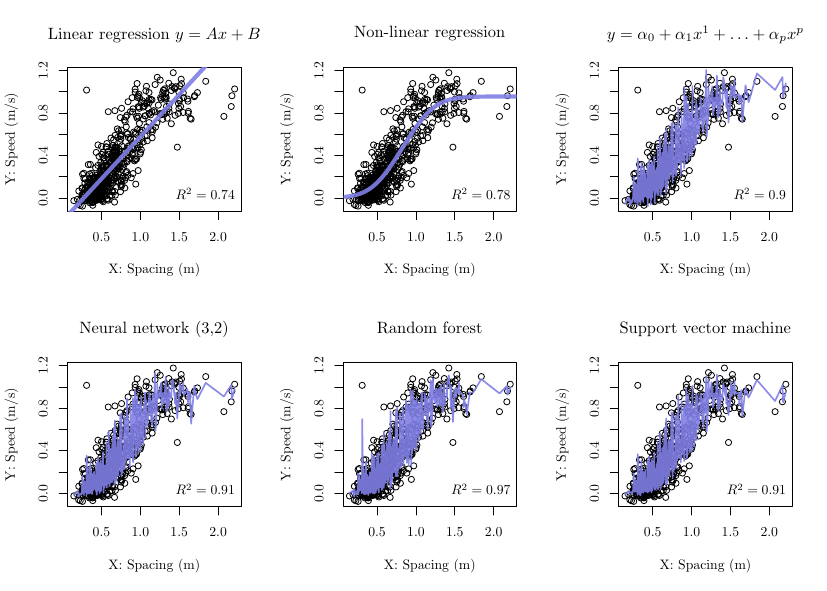}
    \caption{Prediction of the speed with different algorithms. For univariate algorithms (linear and nonlinear regressions in top left and middle panels), the speed is regressed on the distance to the neighbour in front only.}
    \label{fig:PedData2}
\end{figure}

A cross-validated benchmark analysis of different pedestrian speed prediction algorithms is performed\footnote{Benchmarking of pedestrian speed prediction algorithm can be computed online at\\ \href{https://colab.research.google.com/drive/1zGHJQbosu-gZPB8HUaf0GVHSJb_hdcbG?usp=sharing}{\texttt{colab.research.google.com/drive/1zGHJQbosu-gZPB8HUaf0GVHSJb\_hdcbG}}}. 
The testing prediction results and the $R^2$ metric quantifying the goodness of fit are shown in Figure~\ref{fig:PedData2}.
We first consider univariate linear and nonlinear regression before considering multivariate algorithms, including multiple linear regression, a two-layer artificial neural network with three and two neurons respectively, a random forest and a support vector machine. 
Multivariate algorithms clearly outperform univariate algorithms. Among the multivariate algorithms, the random forest performs best, followed by multiple linear regression, artificial neural network and support vector machine which perform similarly.

\end{document}